\documentclass[12pt]{article}

\usepackage{scicite}
\usepackage{amsmath}
\usepackage{amsfonts}
\usepackage{times}
\usepackage{graphicx}
\usepackage{textcomp}
\usepackage{float}
\usepackage{xcolor}
\usepackage{comment}
\usepackage{lineno}

\newenvironment{sciabstract}{%
\begin{quote} \bf}
{\end{quote}}

\title{Depth-Targeted Energy Deposition Deep Inside Scattering Media}

\author
{Nicholas Bender,$^{1}$ Alexey Yamilov,$^{2\ast}$ Arthur Goetschy,$^{3}$\\ Hasan Y\i lmaz,$^{1,4}$ Chia Wei Hsu,$^{5}$ and Hui Cao$^{1\ast}$\\
\\
\normalsize{$^{1}$Department of Applied Physics, Yale University,}\\
\normalsize{New Haven, Connecticut 06520, USA}\\
\normalsize{$^{2}$Physics Department, Missouri University of Science \& Technology,}\\
\normalsize{Rolla, Missouri 65409, USA}\\
\normalsize{$^{3}$ESPCI  Paris,  PSL  University,  CNRS,  Institut  Langevin,}\\
\normalsize{1  rue  Jussieu,  F-75005  Paris,  France}\\
\normalsize{$^{4}$Institute of Materials Science and Nanotechnology,}\\
\normalsize{National Nanotechnology Research Center (UNAM),}\\
\normalsize{Bilkent University, 06800 Ankara, Turkey}\\
\normalsize{$^{5}$Ming Hsieh Department of Electrical and Computer Engineering,}\\
\normalsize{University of Southern California, Los Angeles, California 90089, USA}\\
\\
\normalsize{$^\ast$ Correspondence should be addressed to hui.cao@yale.edu and/or yamilov@mst.edu.}
}

\date{}

\begin{document} 


\baselineskip24pt


\maketitle


\begin{sciabstract}
	
A grand challenge in fundamental physics and practical applications is overcoming wave diffusion to deposit energy into a target region {\it deep} inside a diffusive system. While it is known that coherently controlling the incident wavefront allows diffraction-limited focusing inside a diffusive system, in many applications targets are significantly larger than such a focus and the maximum deliverable energy remains unknown. Here, we introduce the ``deposition matrix'', which maps an input wavefront to its internal field distribution, and theoretically predict the ultimate limitations on energy deposition at any depth. For example, the maximum obtainable energy enhancement occurs at $3/4$ a diffusive system’s thickness: regardless of its scattering strength. Experimentally we measure the deposition matrix and excite its eigenstates to enhance/suppress the energy within an extended target region. Our theoretical analysis reveals that such enhancement/suppression results from both selective transmission eigenchannel excitation and constructive/destructive interference among these channels.

\end{sciabstract}

\section*{Introduction}

Depositing energy into a target region {\it deep inside} an opaque system --by controlling random wave scattering-- is essential in a wide range of applications involving light, microwaves, and acoustic waves~\cite{mosk2012controlling, 2017_Rotter_Gigan_review}: such as deep-tissue imaging ~\cite{yu2015recent, yoon2020deep}, optogenetically controlling neurons~\cite{yoon2015optogenetic, ruan2017deep}, non-invasive ultrasound surgery~\cite{pernot2007vivo}, and optimization of photoelectrochemical processes in strongly-scattering systems~\cite{2016_Liew_ACS_Photon}. The fundamental challenge to overcome in disordered systems is the multiple scattering of waves, which results in a diffusive spread of the wave energy. Controlling the incident wavefront of a coherent beam enables the suppression of wave diffusion; which, has been used to focus light either inside or through a scattering medium~\cite{vellekoop2007focusing, yaqoob2008optical, vellekoop2008demixing, xu2011time, judkewitz2013speckle, horstmeyer2015guidestar, vellekoop2015feedback}. The appropriate incident wavefront can be obtained via the time-reversal principle~\cite{fink2000time}: that the phase conjugate of an output field generated by a point source will focus back to that point \cite{horstmeyer2015guidestar}. Targets in many applications like neurons or early-stage tumors, however, are much larger than an optical-diffraction-limited focal spot and therefore wavelength-scaled light focusing does not corresponds to maximal energy deposition into an extended target. Since the optimal spatial field distribution across the target is not known {\it a priori}, neither time reversal nor phase conjugation can be used to find the optimal incident wavefront. Furthermore, while feedback-based iterative optimization of the input wavefront \cite{vellekoop2015feedback} is efficient at reaching the global maximum when focusing light~\cite{2008_Vellekoop_PRL}; currently, this is not the case for energy delivery into a target of arbitrary size and shape. 

Over the years, various operators and matrices related to physical quantities of interest in disordered systems have been introduced --and their eigenstates studied-- in the search for the global optima of the quantities. Examples include the field transmission matrix~\cite{popoff2010measuring, 2011_Choi_PRB, 2012_Kim_NatPhoton, 2013_Park_PRL, 2014_Popoff, 2014_Aubry_PRL, 2015_Davy_NatCommun,   hsu2017correlation,  yilmaz2019transverse, 2020_Bender_correlations_PRL, HsuPRLBroadband}, the energy density matrix~\cite{cheng2014focusing}, the photoacoustic transmission matrix~\cite{chaigne2014controlling}, the generalized Wigner-Smith operator~\cite{ambichl2017focusing, horodynski2020optimal}, the time-gated reflection matrix~\cite{jeong2018focusing, badon2016smart}, the acousto-optic transmission matrix~\cite{katz2019controlling}, the dwell-time operator~\cite{2019_Durand_PRL}, the distortion matrix~\cite{Aubry_PNAS_2020, Aubry_SciAdv_2020}, and the Fisher information operator~\cite{bouchet2021maximum}. None of them, however, provide the solution for maximal energy deposition in an arbitrary-sized region at an arbitrary depth in a scattering medium.  Furthermore, a general framework for predicting and understanding the ultimate limit on targeted energy delivery into a diffusive system is missing. As such, the following scientifically and technologically important questions remain unanswered, ``How can one systematically find the incident wavefront that optimally deposits energy into a target region of arbitrary size and shape, deep inside a diffusive medium?'' and ``What is the ultimate limit on the energy enhancement in a region?'' 

In this work, we address these questions by performing a comprehensive experimental, numerical and theoretical study. First, we define the deposition matrix $\mathcal{Z}$ which relates input waves to the corresponding regional field distributions at an arbitrary depth within a diffusive system. The largest eigenvalue of $\mathcal{Z}^\dagger\mathcal{Z}$ gives the maximal energy that can be deposited into the designated region, and the associated eigenvector provides the input wavefront. Next, we build a theoretical model which can analytically predict the probability density function of the eigenvalues of $\mathcal{Z}^\dagger\mathcal{Z}$, and demonstrate how energy enhancement depends on the depth of the region and the system parameters such as the transport mean free path $\ell$ and the sample thickness $L$. While the largest possible energy enhancement scales as $L/\ell$, it always occurs at depth $(3/4)L$ in a lossless diffusive medium: independent of the scattering strength. Using a unique on-chip disordered-waveguide platform with an interferometric wavefront shaping setup, we experimentally measure the deposition matrix $\mathcal{Z}$ for regions at different depths inside a diffusive system, and directly excite individual eigenstates to observe their spatial structures across the entire system. Furthermore, we explore the relationship between deposition eigenchannels and transmission eigenchannels; revealing that the regional energy enhancement results from {\it both} selective excitation of high-transmission eigenchannels {\it and} constructive interference between them.

\section*{Experimental Platform and Deposition Matrix}

A schematic of our experimental setup for investigating energy deposition in a diffusive system is presented in Fig.~1. We fabricate two-dimensional (2D) disordered structures, so that we can probe the field anywhere inside the system from the third dimension: the top. We shape the incident wavefront of a laser beam using a spatial light modulator (SLM), and extract the 2D field distribution inside the diffusive system with an interferometric measurement. Our planar samples are optical waveguides engraved into a silicon-on-insulator wafer~\cite{2020_Bender_correlations_PRL}. Light is confined inside the waveguides by reflective photonic-crystal sidewalls. Randomly distributed air holes are etched throughout a designated region in each waveguide to create optical scattering. Light undergoes multiple scattering and diffusive transport within this disordered region because it is much longer than the transport mean free path $\ell$ of the scatterers. A small amount of light is scattered out-of-plane from the holes and interferes with a reference beam. The interference patterns are recorded by a CCD camera, from which the associated field distributions are extracted. Our experimental platform allows for a direct mapping of the incident field to the internal fields at any depth.

Controlling energy deposition inside a disordered system requires introducing the deposition matrix $\mathcal{Z}$ of a target region that can have an arbitrary size, shape, and depth. The matrix relates an orthonormal set of input wavefronts to the corresponding spatial field distributions within the target region (see supplementary section 1). The eigenvalues $\zeta$ of  $\mathcal{Z}^{\dagger}\mathcal{Z}$ give the total energy inside the target region when sending the corresponding eigenvectors into the system: with proper normalization. Therefore, the eigenvector with the highest eigenvalue provides the input wavefront which deposits the most energy into the target region.

As an example case, we consider a target region that is a thin slice inside the disordered waveguide, at depth $z_D$ [see inset of Fig.~2(a)]. The width $W$ of the slice is equal to that of the waveguide, and the slice thickness $\Delta z$ is small enough that the field variation along $z$ (waveguide axis) is negligible. Therefore, only the field distribution along the $y$ axis (waveguide cross-section) needs to be sampled, with $M$ evenly spaced points. For this target region configuration, the elements of the deposition matrix are given by
\begin{equation} 
\mathcal{Z}_{mn}(z_D)\equiv \left(W\Delta z/M \right)^{1/2} E_n(y_m,z_D),
\label{eq:Z_definition1}
\end{equation} 
where $E_n(y_m,z_D)$ is the electric field at position $(y_m,z_D)$ for an incoming wave (of unit flux) in the $n$-th mode of the empty waveguide (input). This definition for the elements of the deposition matrix can easily be generalized to higher dimensions; however, restricting ourselves to a cross-sectional target region facilitates comparison between the deposition matrix and the well-known transmission matrix. Switching to the waveguide-mode basis, the deposition matrix becomes $\mathcal{Z}_{mn}(z_D)=\int_{0}^{W}\chi_m(y) \, E_n(y,z_D) \, dy$, where $\chi_m(y)$ is the normalized transverse profile of the $m$-th mode of a homogeneous waveguide with a refractive index equal to the average index of the disordered region. Note that the waveguide modes include both propagating modes and evanescent modes. If the evanescent waves are negligible, only the propagating modes are kept and normalized by their propagation speed $v_m$, we get 
\begin{equation} 
\mathcal{Z}_{mn}(z_D)= \sqrt{v_m} \, \int_{0}^{W}\chi_m(y) \, E_n(y,z_D) \, dy.
\label{eq:Z_definition2}
\end{equation}  
In this form, the deposition matrix naturally reduces to the transmission matrix at the end of the disordered region $z_D = L$. In our disordered waveguides, the deposition matrices defined by Eqs.~(1) and (2) have nearly identical eigenvalues and eigenvectors for most depths except when very close to the exit surface $z_D = L$. More details are given in supplementary section 1.

\section*{Numerical Simulation and Analytical Model}

To reveal the full potential of the deposition matrix (DM) for energy deposition inside disordered systems, we first carry out numerical simulations of wave propagation in 2D disordered waveguides using the Kwant software package~\cite{2014_Groth_Kwant} (see supplementary section 1.1 for details). For comparison with the transmission matrix (TM), we adopt the DM defined by Eq.~(2) and calculate its eigenvalues $\zeta$ for a thin slice at different depths $z_D$ inside a lossless disordered waveguide. The probability density function (PDF) $P(\zeta)$, shown in Fig.~2(a), is very different from the celebrated bimodal PDF of transmission eigenvalues $P(\tau)$~\cite{1997_Beenakker}. At depths $z_D<L$, $P(\zeta)$ has a single peak at $\zeta = 0$, but it develops a second peak at $\zeta = 1$ near the exit surface $L-z_d < \ell$ (shaded area). We normalize the eigenvalues $\zeta$ by their mean $\left<\zeta\right>$, which represents the typical energy within the slice at depth $z_D$ under random illumination conditions. Despite the lack of a peak at the maximum eigenvalue $\zeta_\text{\rm max}$, for most depths $P(\zeta/\left<\zeta\right>)$ has a long tail extending beyond the range of  $P(\tau/\left<\tau\right>)$ . Consequently, the maximal enhancement of energy inside the diffusive system, given by $\zeta_\text{\rm max} / \left<\zeta\right>$, is noticeably larger than the maximum enhancement of the transmission $\tau_\text{\rm max} / \left<\tau\right>$ for open channels ($\tau_\text{\rm max} = 1$, $\left<\tau\right> \sim \ell /L \ll 1$). 

 To interpret these results quantitatively, we develop an analytical model for the PDF of the deposition eigenvalues $P(\zeta)$. The DM $\mathcal{Z}(z_D)$ cannot be treated as a random matrix with uncorrelated matrix elements, because the eigenvalue PDF in Fig.~2(a) drastically differs from the Marchenko-Pastur law~\cite{1967_Marchenko}. In particular, the latter predicts $\left<\zeta_\text{\rm max} \right>/ \left<\zeta\right>=4$, whereas significantly larger values are obtained at almost all depths, indicating that correlations between elements of $\mathcal{Z}(z_D)$ are beneficial for energy deposition. Since the DM and the TM coincide at the exit, we build a model that captures the continuous evolution from $P(\zeta)$ at $z_D<L$ to the bimodal PDF at $z_D = L$. This is realized by using a filtered random matrix (FRM) ensemble as initially introduced in Ref.~\cite{2013_Goetschy}. This theory amounts to assuming that  $\mathcal{Z}(z_D)$ has the same spectrum as a filtered matrix drawn from a larger virtual TM (see supplementary section 2). The advantage of this approach is that the full PDF $P(\zeta)$ can be inferred from the first two moments $\left<\zeta\right>$ and $\left<\zeta^2\right>$. Here we use the numerical values of these two moments as input parameters of the model. The good agreement between the numerical PDF and the FRM prediction in Fig.~2(a) validates our ansatz.

Combining the FRM model with analytic predictions for the first two moments of $P(\zeta)$, we get analytical expressions for the full PDF as well as the maximal enhancement. The first moment decays linearly with depth,  $\left<\zeta(z_D)\right>\simeq 2(1-\left<\tau \right>)(1-z_D/L)+\left<\tau \right>$, as given by diffusion theory~\cite{1999_VanRossum}. The second moment is given by the variance $\text{Var}[\zeta(z_D)]$, which is related to the fluctuation of the cross-section integrated intensity at depth $z_D$ generated by random wavefront illumination~\cite{2014_Popoff}: $\text{Var}[\zeta(z_D)]\simeq \left<\zeta\right>^2 [1+N \, C_2(z_D)]$. In this expression, $N$ is the number of waveguide modes in the disordered region and $C_2(z_D)$ stands for long-range contributions to the spatial intensity correlation function: whose analytic expressions are given in \cite{1989_Pnini, 2014_Sarma}. Combining these with the FRM model, in the limit $N\gg 1$, we predict a finite support for $P(\zeta)$ and thus a maximal energy enhancement given by the upper edge of $P(\zeta)$. Fig.~2(b) shows a quantitative agreement between this prediction and $\left<\zeta_\text{\rm max} \right>/\left<\zeta \right>$ evaluated numerically for disordered waveguides of different sizes and scattering strengths: without any adjustable parameter. The FRM predicts that $\left<\zeta_\text{\rm max} \right>/\left<\zeta \right>$ depends only on $C_2(z_D)$ for most depths $z_D$, confirming the crucial role of mesoscopic correlations in enhancing energy deposition. The general expression for the energy enhancement is derived and presented in the supplementary (section 2), below we present a simplified form in the limit of $L \gg \ell$: 
\begin{equation}
\frac{\left<\zeta_\text{\rm max} (z_D) \right>}{\left<\zeta (z_D) \right>}\simeq \frac{3N \, C_2(z_D)}{2} \simeq \frac{ 3(z_D/L)-2(z_D/L)^2} {\left<\tau\right>}.
\label{eq:zeta_max}
\end{equation} 
Two conclusions can be drawn from this result. First, the maximal energy enhancement is inversely proportional to $\left<\tau\right>$ and thus grows linearly with $L/\ell$. In particular, it is independent of the width $W$ of the disordered waveguide as long as the dimensionless conductance $g=N \left<\tau \right>$ is sufficiently large. Second, apart from  $\left<\tau\right>$, the energy enhancement depends on the reduced depth $z_D/L$ only; reaching a maximal value of $9/8\langle\tau\rangle \sim L/\ell \gg 1$ at $z_D^{\rm (max)} /L \sim 3/4$. This result holds for different transport mean free paths, as confirmed in Fig. 2(b). Hence, the largest enhancement is {\textit not} obtained at the output surface, but rather deep inside the diffusive medium at depth $3L/4$: independent of $\ell$.

\section*{Measurement of Deposition Eigenchannels}

We experimentally measure different deposition matrices in disordered waveguides like the one shown in Fig.~3(a). The disordered region of each waveguide is $L =50$ \textmu m long and $W = 15$ \textmu m wide. The transport mean free path at the optical wavelength $\lambda = 1.55$ \textmu m is $\ell = 3.2$ \textmu m. The out-of-plane scattering loss is not negligible, however as in \cite{2014_Yamilov_PRL}, it can be modeled through an effective diffusive dissipation length $\xi_a$ = 28 \textmu m. We construct the deposition matrices associated with four target regions inside the disordered waveguide: each is 10 \textmu m $\times$ 10 \textmu m. They are centered at depths $z_D$ = 10, 20, 30, 40 \textmu m.      

We use a SLM to modulate the monochromatic laser beam incident on the waveguide, and measure the field distribution within each target region (for details see supplementary section 3.3). From the data, we extract the DM and perform a singular value decomposition to obtain the deposition eigenchannels’ input vectors. These vectors are the eigenvectors of $\mathcal{Z}^\dagger\mathcal{Z}$; each is sorted by its corresponding eigenvalue, from high to low, and labeled by an index $\alpha$. We sequentially shape the incident wavefront into each of the eigenvectors, thereby exciting one eigenchannel at a time, and record the 2D intensity distribution over the entire disordered waveguide. The cross-section integrated intensity $I^{(D)}_\alpha(z)$ depicts the depth profile of every eigenchannel. We repeat this measurement for multiple disorder realizations --generated at multiple wavelengths and with different hole configurations-- and ensemble average the spatial profiles of the eigenchannels with the same index $\alpha$.

In Fig.~3(b,c), we show the depth profiles of example eigenchannels with enhanced or suppressed energy deposition, for two different target regions. Both strong energy enhancement and suppression are observed experimentally in the target region --when compared to the average depth profile $\left< I(z) \right>$ of random illumination patterns-- and reproduced numerically. Simultaneously the energy outside the target region is enhanced or suppressed, reflecting the non-local effects in the energy deposition. Quantitatively, we compute the energy enhancement factor in the target region
$\eta_t = \int_{z\subset R} I_{\alpha}(z) dz /\int_{z\subset R} \langle I (z) \rangle dz $, and in the surrounding area $\eta_s = \int_{z\not\subset R} I_{\alpha}(z) dz /\int_{z\not\subset R} \langle I (z) \rangle dz $. Figure~3(d) shows that $\eta_t$ increases with depth $z_D$, while $\eta_s$ remains nearly constant. The depth variation of the regional enhancement $\eta_t(z_D)$ is captured by the long-range correlation function $C_{2}(z_D)$, in agreement with our theoretical model. Due to the presence of loss in the diffusive waveguide, the depth of the maximal energy enhancement --which coincides with the maximum of $C_{2}(z_D)$-- is slightly shifted from $z_D=(3/4)L$ towards the output end. Figure~3(e) shows that the suppression of energy within the target region gets stronger for larger depths, but the suppression in the surrounding area is independent of depth.   

\section*{Two mechanisms for energy deposition}

To gain physical insight into the formation of deposition eigenchannels and how they enhance or suppress energy within local regions {\it inside} a diffusive system, we decompose them into the transmission eigenchannels, whose spatial profiles have been studied extensively~\cite{2011_Choi_PRB, 2015_Davy_NatCommun, 2016_Sarma_Open_Channels, 2016_Ojambati_OptExpress, 2017_Koirala_Inverse_Design, 2018_Hong_Optica, 2019_Tian_PRB, 2020_Bender_correlations_PRL}. At the entrance of the system $z = 0$, the transmission eigenvectors form a complete basis, and the input wavefront of a deposition eigenchannel can be expressed as a linear superposition of the transmission eigenchannels. The linear mapping from the incident field to the internal field carries the decomposition to the entire field distribution inside the disordered waveguide: $E^{(D)}_\alpha(y,z) = \sum_{\beta=1}^N d_{\alpha \beta} E^{(T)}_\beta(y,z)$. In this expression $E^{(D)}_\alpha(y,z)$ [$E^{(T)}_\beta(y,z)$] denotes the field distribution of the $\alpha$-th deposition ($\beta$-th transmission) eigenchannel and $N$ is the number of transmission eigenchannels (equal to the number of propagating modes in the input waveguide). The depth profile of a deposition channel, given by the cross-section integrated intensity  $I^{(D)}_\alpha(z) = \int_0^W |E^{(D)}_\alpha(y,z)|^2 \, dy$, consists of two terms:
\begin{equation}
\begin{split}
I^{(D)}_\alpha(z) & = I_\alpha^{(i)}(z) + I_\alpha^{(c)}(z)  \\ 
& = \sum_{\beta=1}^N |d_{\alpha \beta}|^2 \, I^{(T)}_\beta(z) + \sum_{\beta \neq \beta'} d_{\alpha \beta} \, d^*_{\alpha \beta'} \, I^{(T)}_{\beta \, \beta'}(z). \nonumber 
\end{split}
\end{equation}
The first term $I_\alpha^{(i)}(z)$ is an incoherent sum of the constituent transmission eigenchannel depth profiles, $I^{(T)}_\beta(z) = \int_0^W |E^{(T)}_\beta(y,z)|^2 \, dy$, studied in~\cite{2011_Choi_PRB, 2015_Davy_NatCommun, 2016_Sarma_Open_Channels, 2016_Ojambati_OptExpress, 2017_Koirala_Inverse_Design, 2018_Hong_Optica, 2019_Tian_PRB, 2020_Bender_correlations_PRL}. The second term $I_\alpha^{(c)}(z)$ is the result of interference between different transmission eigenchannels inside the diffusive waveguide, which we observe for the first time. Although the transmission eigenchannels are orthogonal at $z=0$ and $z=L$, this is not the case inside: $I^{(T)}_{\beta \, \beta'}(z) = \int_0^W  E^{(T)}_\beta(y,z) \, E^{(T)*}_{\beta'}(y,z) \, dy \neq 0$ for $0<z< L$.

To find how much these two terms contribute to the energy enhancement, we numerically decompose the maximal energy deposition eigenchannels ($\alpha = 1$) for the four target regions inside our disordered waveguide. As shown in  Fig.~\ref{fig:composition_of_DE}(a), each is composed of multiple high-transmission eigenchannels (higher transmission corresponds to lower index $\beta$). With increasing depth $z_D$, the number of constituent transmission eigenchannels decreases, and the maximal decomposition coefficient $|d_{\alpha \beta}|^2$ shifts to $\beta = 1$ (the highest-transmission eigenchannel). Figure~\ref{fig:composition_of_DE}(b) shows the incoherent contribution $I_1^{(i)}(z)$ and coherent contribution $I_1^{(c)}(z)$ to energy deposition in the target region. When the target region is located at a shallower depth, more transmission eigenchannels participate in constructing the deposition eigenchannel, and their constructive interference plays an important role in enhancing energy deposition in the target region. As the number of participating transmission eigenchannels becomes progressively smaller with increasing depth, the interference effect is weakened and the incoherent contribution from selective excitation of transmission eigenchannels becomes dominant. 

We also investigate the deposition eigenchannels that reduce energy within the target regions. As shown in Fig.~\ref{fig:composition_of_DE}(c), the $\alpha = 25$ deposition eigenchannels consist of multiple transmission eigenchannels with indices $\beta$ close to 25. The suppression of energy within the target region results from selective excitation of lower-transmission eigenchannels and their destructive interference [see Fig.~\ref{fig:composition_of_DE}(d)]. The deeper the target region, the lower the number of constituent transmission eigenchannels, the weaker their destructive interference effect. Thanks to the destructive interference, the total transmission is less than the energy inside the target region. Thus, when sending light through a diffusive system it is possible to avoid certain regions inside.   

\section*{Discussion and conclusions}

In conclusion, we have delineated the fundamental limits on depositing energy into a finite region, located at any depth, inside a diffusive system. In contrast to the bimodal distribution of transmission eigenvalues, the PDF of deposition eigenvalues $P(\zeta)$ has only one peak at $\zeta = 0$ and a long tail for most depths: $\zeta / \left< \zeta \right> \gg 1$. Our theoretical model, based on a filtered random matrix ensemble, can analytically predict $P(\zeta)$ for regions anywhere inside a diffusive medium. The long-range correlations present in the intensity of the field, induced by the multiple scattering of light and characterized by $C_{2}(z_{D})$, facilitate optical energy deposition. In a diffusive waveguide of length $L$ much larger than the transport mean free path $\ell$, the largest possible energy enhancement $\left<\zeta_\text{\rm max} \right>/\left<\zeta \right>$ at a depth $z_D$ depends only on two parameters: $L/\ell$ and $z_D/L$. With increasing depth $z_D$, $\left<\zeta_\text{\rm max} \right>/\left<\zeta \right>$ rises and reaches a global maximum $\sim L/\ell$ at $z_D^{\rm (max)}/L \sim 3/4$. Because $z_D^{\rm (max)}$ is dependent on $L$ and independent of $\ell$, when $L \gg \ell$, the depth of the maximal enhancement is deep inside the sample rather than near the front or back surfaces. Although our experimental and numerical studies are conducted on 2D systems, the above scaling results follow from filtered matrix theory, e.g. Eq.~\ref{eq:zeta_max}, which also applies in three dimensions.

Additionally, we discovered the relationship between deposition eigenchannels and transmission eigenchannels. We found that it is impossible to construct the intensity profile of a deposition eigenchannel from the intensity profiles of the transmission eigenchannels alone. Constructive or destructive interference between transmission eigenchannels inside the disordered system plays a prominent role in enhancing or suppressing energy within the target region. Therefore, our analysis reveals two distinct mechanisms for energy deposition: selective excitation of transmission eigenchannels and interference between them. Their contributions are characterized by the amplitudes and phases of the coefficients obtained when decomposing a deposition eigenchannel into a summation of transmission eigenchannels.   

Although our studies are conducted on planar waveguides with narrow widths and transverse confinement, we believe the conclusions can be extended to wide slabs with open boundaries and to volumetric diffusive systems. They are also applicable to other types of waves such as microwaves and acoustic waves. Targeted energy delivery opens the door to numerous applications, e.g., optogenetic control of cells, photothermal therapy, as well as probing and manipulating photoelectrochemical processes deep inside nominally opaque media.

\bibliography{Citations}

\begin{thebibliography}{10}

\bibitem{mosk2012controlling}
A.~P. Mosk, A.~Lagendijk, G.~Lerosey, M.~Fink, Controlling waves in space and
  time for imaging and focusing in complex media, {\it Nat. Photon.\/} {\bf 6},
  283 (2012).

\bibitem{2017_Rotter_Gigan_review}
S.~Rotter, S.~Gigan, Light fields in complex media: mesoscopic scattering meets
  wave control, {\it Rev. Mod. Phys.\/} {\bf 89}, 015005 (2017).

\bibitem{yu2015recent}
H.~Yu, {\it et~al.\/}, Recent advances in wavefront shaping techniques for
  biomedical applications, {\it Curr. Appl. Phys.\/} {\bf 15}, 632 (2015).

\bibitem{yoon2020deep}
S.~Yoon, {\it et~al.\/}, Deep optical imaging within complex scattering media,
  {\it Nat. Rev. Phys.\/} {\bf 2}, 141 (2020).

\bibitem{yoon2015optogenetic}
J.~Yoon, {\it et~al.\/}, Optogenetic control of cell signaling pathway through
  scattering skull using wavefront shaping, {\it Sci. Rep.\/} {\bf 5}, 1
  (2015).

\bibitem{ruan2017deep}
H.~Ruan, {\it et~al.\/}, Deep tissue optical focusing and optogenetic
  modulation with time-reversed ultrasonically encoded light, {\it Sci. Adv.\/}
  {\bf 3}, eaao5520 (2017).

\bibitem{pernot2007vivo}
M.~Pernot, {\it et~al.\/}, In vivo transcranial brain surgery with an
  ultrasonic time reversal mirror, {\it J. Neurosurg.\/} {\bf 106}, 1061
  (2007).

\bibitem{2016_Liew_ACS_Photon}
S.~F. Liew, {\it et~al.\/}, Coherent control of photocurrent in a strongly
  scattering photoelectrochemical system, {\it ACS Photon.\/} {\bf 3}, 449
  (2016).

\bibitem{vellekoop2007focusing}
I.~M. Vellekoop, A.~Mosk, Focusing coherent light through opaque strongly
  scattering media, {\it Opt. Lett.\/} {\bf 32}, 2309 (2007).

\bibitem{yaqoob2008optical}
Z.~Yaqoob, D.~Psaltis, M.~S. Feld, C.~Yang, Optical phase conjugation for
  turbidity suppression in biological samples, {\it Nat. Photon.\/} {\bf 2},
  110 (2008).

\bibitem{vellekoop2008demixing}
I.~M. Vellekoop, E.~Van~Putten, A.~Lagendijk, A.~Mosk, Demixing light paths
  inside disordered metamaterials, {\it Opt. Express\/} {\bf 16}, 67 (2008).

\bibitem{xu2011time}
X.~Xu, H.~Liu, L.~V. Wang, Time-reversed ultrasonically encoded optical
  focusing into scattering media, {\it Nat. Photon.\/} {\bf 5}, 154 (2011).

\bibitem{judkewitz2013speckle}
B.~Judkewitz, Y.~M. Wang, R.~Horstmeyer, A.~Mathy, C.~Yang, Speckle-scale
  focusing in the diffusive regime with time reversal of variance-encoded light
  {(TROVE)}, {\it Nat. Photon.\/} {\bf 7}, 300 (2013).

\bibitem{horstmeyer2015guidestar}
R.~Horstmeyer, H.~Ruan, C.~Yang, Guidestar-assisted wavefront-shaping methods
  for focusing light into biological tissue, {\it Nat. Photon.\/} {\bf 9}, 563
  (2015).

\bibitem{vellekoop2015feedback}
I.~M. Vellekoop, Feedback-based wavefront shaping, {\it Opt. Express\/} {\bf
  23}, 12189 (2015).

\bibitem{fink2000time}
M.~Fink, {\it et~al.\/}, Time-reversed acoustics, {\it Rep. Prog. Phys.\/} {\bf
  63}, 1933 (2000).

\bibitem{2008_Vellekoop_PRL}
I.~M. Vellekoop, A.~P. Mosk, Universal optimal transmission of light through
  disordered materials, {\it Phys. Rev. Lett.\/} {\bf 101}, 120601 (2008).

\bibitem{popoff2010measuring}
S.~Popoff, {\it et~al.\/}, Measuring the transmission matrix in optics: an
  approach to the study and control of light propagation in disordered media,
  {\it Phys. Rev. Lett.\/} {\bf 104}, 100601 (2010).

\bibitem{2011_Choi_PRB}
W.~Choi, A.~P. Mosk, Q.-H. Park, W.~Choi, Transmission eigenchannels in a
  disordered medium, {\it Phys. Rev. B\/} {\bf 83}, 134207 (2011).

\bibitem{2012_Kim_NatPhoton}
M.~Kim, {\it et~al.\/}, Maximal energy transport through disordered media with
  the implementation of transmission eigenchannels, {\it Nat. Photon.\/} {\bf
  6}, 581 (2012).

\bibitem{2013_Park_PRL}
H.~Yu, {\it et~al.\/}, Measuring large optical transmission matrices of
  disordered media, {\it Phys. Rev. Lett.\/} {\bf 111}, 153902 (2013).

\bibitem{2014_Popoff}
S.~M. Popoff, A.~Goetschy, S.~F. Liew, A.~D. Stone, H.~Cao, Coherent control of
  total transmission of light through disordered media, {\it Phys. Rev.
  Lett.\/} {\bf 112}, 133903 (2014).

\bibitem{2014_Aubry_PRL}
B.~G\'erardin, J.~Laurent, A.~Derode, C.~Prada, A.~Aubry, Full transmission and
  reflection of waves propagating through a maze of disorder, {\it Phys. Rev.
  Lett.\/} {\bf 113}, 173901 (2014).

\bibitem{2015_Davy_NatCommun}
M.~Davy, Z.~Shi, J.~Park, C.~Tian, A.~Z. Genack, Universal structure of
  transmission eigenchannels inside opaque media, {\it Nat. Commun.\/} {\bf 6},
  6893 (2015).

\bibitem{hsu2017correlation}
C.~W. Hsu, S.~F. Liew, A.~Goetschy, H.~Cao, A.~D. Stone, Correlation-enhanced
  control of wave focusing in disordered media, {\it Nat. Phys.\/} {\bf 13},
  497 (2017).

\bibitem{yilmaz2019transverse}
H.~Y{\i}lmaz, C.~W. Hsu, A.~Yamilov, H.~Cao, Transverse localization of
  transmission eigenchannels, {\it Nat. Photon.\/} {\bf 13}, 352 (2019).

\bibitem{2020_Bender_correlations_PRL}
N.~Bender, A.~Yamilov, H.~Y{\i}lmaz, H.~Cao, Fluctuations and correlations of
  transmission eigenchannels in diffusive media, {\it Phys. Rev. Lett.\/} {\bf
  125}, 165901 (2020).

\bibitem{HsuPRLBroadband}
C.~W. Hsu, A.~Goetschy, Y.~Bromberg, A.~D. Stone, H.~Cao, Broadband coherent
  enhancement of transmission and absorption in disordered media, {\it Phys.
  Rev. Lett.\/} {\bf 115}, 223901 (2015).

\bibitem{cheng2014focusing}
X.~Cheng, A.~Z. Genack, Focusing and energy deposition inside random media,
  {\it Opt. Lett.\/} {\bf 39}, 6324 (2014).

\bibitem{chaigne2014controlling}
T.~Chaigne, {\it et~al.\/}, Controlling light in scattering media
  non-invasively using the photoacoustic transmission matrix, {\it Nat.
  Photon.\/} {\bf 8}, 58 (2014).

\bibitem{ambichl2017focusing}
P.~Ambichl, {\it et~al.\/}, Focusing inside disordered media with the
  generalized {Wigner-Smith} operator, {\it Phys. Rev. Lett.\/} {\bf 119},
  033903 (2017).

\bibitem{horodynski2020optimal}
M.~Horodynski, {\it et~al.\/}, Optimal wave fields for micromanipulation in
  complex scattering environments, {\it Nat. Photon.\/} {\bf 14}, 149 (2020).

\bibitem{jeong2018focusing}
S.~Jeong, {\it et~al.\/}, Focusing of light energy inside a scattering medium
  by controlling the time-gated multiple light scattering, {\it Nat. Photon.\/}
  {\bf 12}, 277 (2018).

\bibitem{badon2016smart}
A.~Badon, {\it et~al.\/}, Smart optical coherence tomography for ultra-deep
  imaging through highly scattering media, {\it Sci. Adv.\/} {\bf 2}, e1600370
  (2016).

\bibitem{katz2019controlling}
O.~Katz, F.~Ramaz, S.~Gigan, M.~Fink, Controlling light in complex media beyond
  the acoustic diffraction-limit using the acousto-optic transmission matrix,
  {\it Nat. Commun.\/} {\bf 10}, 1 (2019).

\bibitem{2019_Durand_PRL}
M.~Durand, S.~M. Popoff, R.~Carminati, A.~Goetschy, Optimizing light storage in
  scattering media with the dwell-time operator, {\it Phys. Rev. Lett.\/} {\bf
  123}, 243901 (2019).

\bibitem{Aubry_PNAS_2020}
W.~Lambert, L.~A. Cobus, T.~Frappart, M.~Fink, A.~Aubry, Distortion matrix
  approach for ultrasound imaging of random scattering media, {\it Proc. Natl.
  Acad. Sci. U.S.A\/} {\bf 117}, 14645 (2020).

\bibitem{Aubry_SciAdv_2020}
A.~Badon, {\it et~al.\/}, Distortion matrix concept for deep optical imaging in
  scattering media, {\it Sci. Adv.\/} {\bf 6}, eaay7170 (2020).

\bibitem{bouchet2021maximum}
D.~Bouchet, S.~Rotter, A.~P. Mosk, Maximum information states for coherent
  scattering measurements, {\it Nat. Phys.\/} pp. 1--5 (2021).

\bibitem{2014_Groth_Kwant}
C.~W. Groth, M.~Wimmer, A.~R. Akhmerov, X.~Waintal, Kwant: A software package
  for quantum transport, {\it New J. Phys.\/} {\bf 16} (2014).

\bibitem{1997_Beenakker}
C.~W. Beenakker, Random-matrix theory of quantum transport, {\it Rev. Mod.
  Phys.\/} {\bf 69}, 731 (1997).

\bibitem{1967_Marchenko}
V.~A. Marchenko, L.~A. Pastur, Distribution of eigenvalues for some sets of
  random matrices, {\it Math. USSR Sbornik\/} {\bf 1}, 457 (1967).

\bibitem{2013_Goetschy}
A.~Goetschy, A.~D. Stone, Filtering random matrices: the effect of incomplete
  channel control in multiple scattering, {\it Phys. Rev. Lett.\/} {\bf 111},
  063901 (2013).

\bibitem{1999_VanRossum}
M.~C.~W. van Rossum, T.~M. Nieuwenhuizen, Multiple scattering of classical
  waves: microscopy, mesoscopy, and diffusion, {\it Rev. Mod. Phys.\/} {\bf
  71}, 313 (1999).

\bibitem{1989_Pnini}
R.~Pnini, B.~Shapiro, Fluctuations in transmission of waves through disordered
  slabs, {\it Phys. Rev. B\/} {\bf 39}, 6986 (1989).

\bibitem{2014_Sarma}
R.~Sarma, A.~Yamilov, P.~Neupane, B.~Shapiro, H.~Cao, Probing long-range
  intensity correlations inside disordered photonic nanostructures, {\it Phys.
  Rev. B\/} {\bf 90}, 014203 (2014).

\bibitem{2014_Yamilov_PRL}
A.~G. Yamilov, {\it et~al.\/}, Position-dependent diffusion of light in
  disordered waveguides, {\it Phys. Rev. Lett.\/} {\bf 112}, 023904 (2014).

\bibitem{2016_Sarma_Open_Channels}
R.~Sarma, A.~Yamilov, S.~Petrenko, Y.~Bromberg, H.~Cao, Control of energy
  density inside a disordered medium by coupling to open or closed channels,
  {\it Phys.~Rev.~Lett.\/} {\bf 117} (2016).

\bibitem{2016_Ojambati_OptExpress}
O.~S. Ojambati, A.~P. Mosk, I.~M. Vellekoop, A.~Lagendijk, W.~L. Vos, Mapping
  the energy density of shaped waves in scattering media onto a complete set of
  diffusion modes, {\it Opt. Express\/} {\bf 24}, 18525 (2016).

\bibitem{2017_Koirala_Inverse_Design}
M.~Koirala, R.~Sarma, H.~Cao, A.~Yamilov, Inverse design of perfectly
  transmitting eigenchannels in scattering media, {\it Phys. Rev. B\/} {\bf 96}
  (2017).

\bibitem{2018_Hong_Optica}
P.~Hong, O.~S. Ojambati, A.~Lagendijk, A.~P. Mosk, W.~L. Vos, Three-dimensional
  spatially resolved optical energy density enhanced by wavefront shaping, {\it
  Optica\/} {\bf 5}, 844 (2018).

\bibitem{2019_Tian_PRB}
P.~Fang, {\it et~al.\/}, Universality of eigenchannel structures in dimensional
  crossover, {\it Phys. Rev. B\/} {\bf 99}, 094202 (2019).

\bibitem{1981_Fisher}
D.~S. Fisher, P.~A. Lee, Relation between conductivity and transmission matrix,
  {\it Phys.~Rev.~B\/} {\bf 23}, 6851 (1981).

\bibitem{1990_Mello}
P.~A. Mello, Averages on the unitary group and applications to the problem of
  disordered conductors, {\it J. Phys. A\/} {\bf 23}, 4061 (1990).

\bibitem{CCCL}
N.~Bender, H.~Yılmaz, Y.~Bromberg, H.~Cao, Creating and controlling complex
  light, {\it APL Photon.\/} {\bf 4}, 110806 (2019).

\end{thebibliography}

\bibliographystyle{Science}

\section*{Acknowledgments}
H.C. thanks Azriel Genack for stimulating discussions. 
{\bf Funding}: This work is supported partly by the Office of Naval Research (ONR) under Grant No. N00014-20-1-2197, and by the National Science Foundation under Grant Nos. DMR-1905465, DMR-1905442, OAC-1919789. 
{\bf Author contributions}: N.B. conducted the experiments and analyzed the data; A.Y. performed the numerical simulations. A.G. developed the analytical model. H.Y. participated in the experimental study. C.W.H. contributed to the theoretical analysis. H.C. initiated the project and supervised the research. All authors contributed to the manuscript preparation. {\bf Competing interests}: The authors
declare no competing financial interests. 
{\bf Data and materials availability}: All data needed to evaluate the conclusions in the paper are present in the paper or the supplementary materials. 

\newpage
\begin{figure}[H]
\begin{center}
\includegraphics[width=\textwidth]{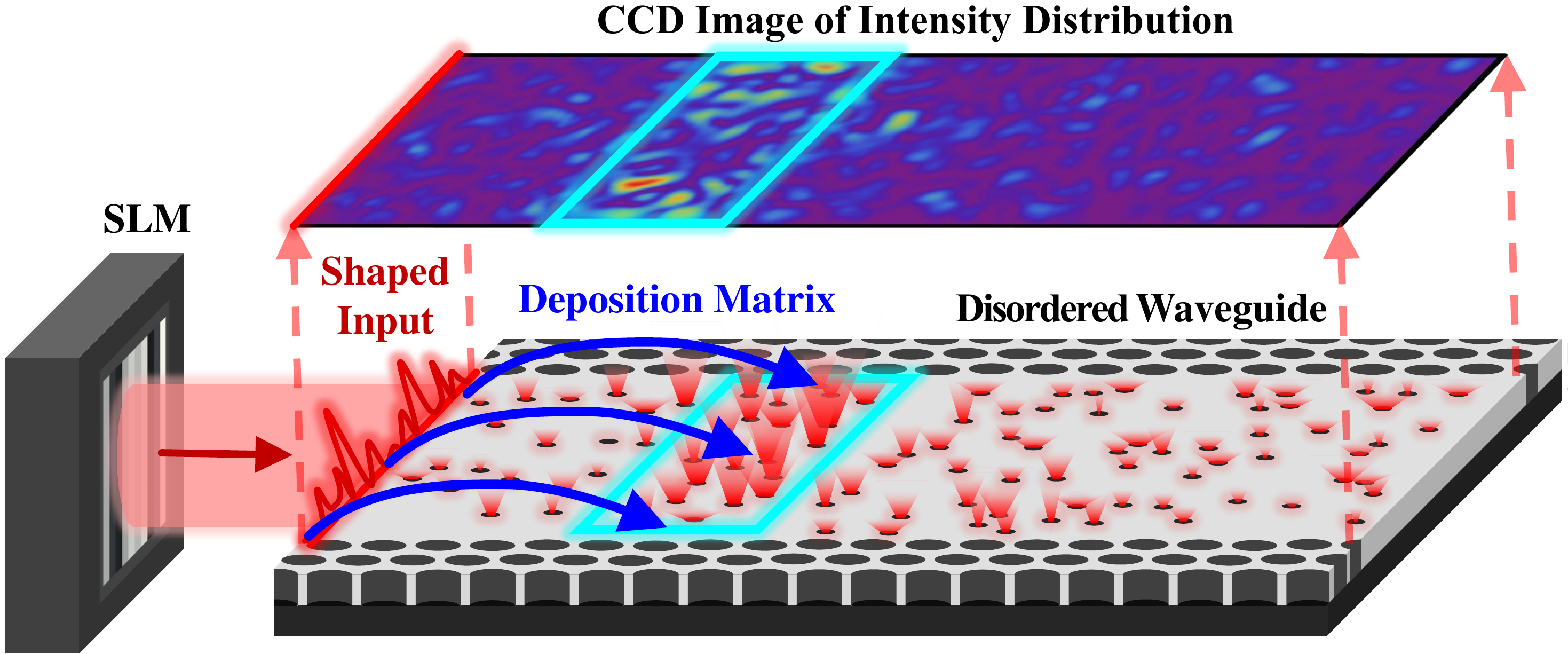}
\caption{{\bf Schematic experimental platform for investigating energy deposition in a diffusive system.} A spatial light modulator (SLM) shapes the incident wavefront of a monochromatic laser beam, and the field distribution inside a two-dimensional disordered waveguide is probed from above. This setup allows measurement of the deposition matrix that relates the incoming field pattern to the spatial field distribution inside a target region (marked by the cyan box). Selective coupling of light into the deposition eigenchannels can enhance or suppress energy inside the target region, as confirmed by the CCD camera image of the spatial intensity distribution.
}
\end{center}
\end{figure}

\newpage
\begin{figure}[H]
\begin{center}
\includegraphics[width=4in]{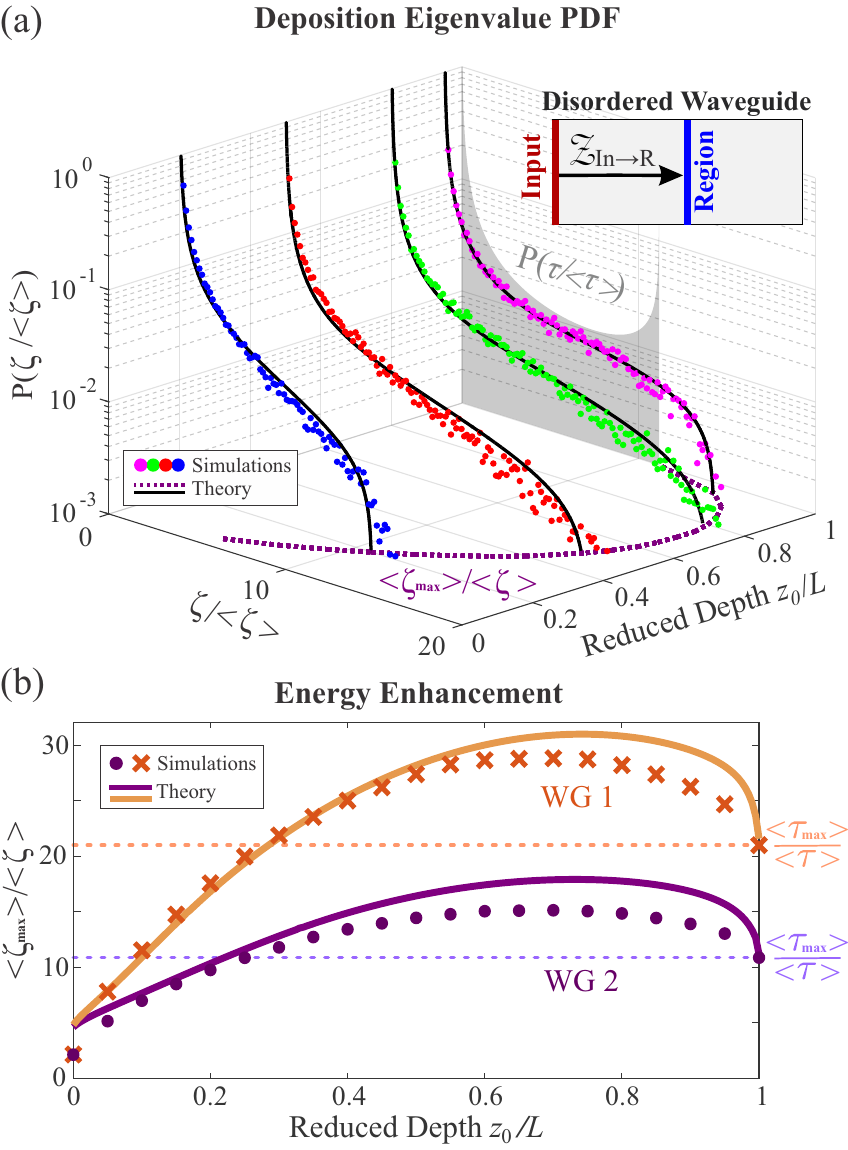}
\caption{ 
{\bf Numerical simulation and analytic prediction of deposition eigenvalues.}
(a) Probability density function of normalized deposition eigenvalues $\zeta/ \left< \zeta \right>$ for a thin slice at varying depths $z_D$ inside a diffusive waveguide (see inset). Analytical FRM predictions (solid lines) agree with numerical simulations (dots) averaged over $1000$ disorder configurations. For most depths, $P(\zeta/ \left< \zeta \right>)$ is very different from the bimodal distribution of the transmission eigenvalues $P(\tau/ \left< \tau \right>)$, although it converges to bimodal at the end (shaded area at $z_D/L = 1$). The theoretical prediction for the upper edge of $P(\zeta)$, which sets the limit for energy enhancement $\left<\zeta_\text{\rm max} \right>/\left<\zeta \right>$, is marked by dashed purple line in the horizontal plane. (b) Energy enhancement in two diffusive waveguides (WG1, WG2), given by the ratio of the largest ensemble-averaged deposition-eigenvalue $\left<\zeta_\text{\rm max} \right>$ over the mean eigenvalue $\left<\zeta \right>$, increases with depth $z_D$ and reaches its maximum at $z_D/L \sim 3/4$. Analytical predictions for the upper edge of $P(\zeta/ \left< \zeta \right>)$ (solid lines) are compared to numerical data (symbols). The energy enhancement $\left<\tau_\text{\rm max} \right>/\left<\tau \right>$ exceeds the transmission enhancement $\left<\zeta_\text{\rm max} \right>/\left<\zeta \right>$ (horizontal dotted line) at most depths. In (a), the waveguide (WG1) has a length $L=50\, \mu$m,  width $W=15\, \mu$m, and transport mean free path $\ell=3.3 \, \mu$m. (b) includes a second waveguide (WG2) of $L=50\, \mu$m,  $W=30\, \mu$m, and $\ell= 1.6 \, \mu$m. 
}
\end{center}
\end{figure}

\newpage
\begin{figure}[H]
\begin{center}
\includegraphics[width=4in]{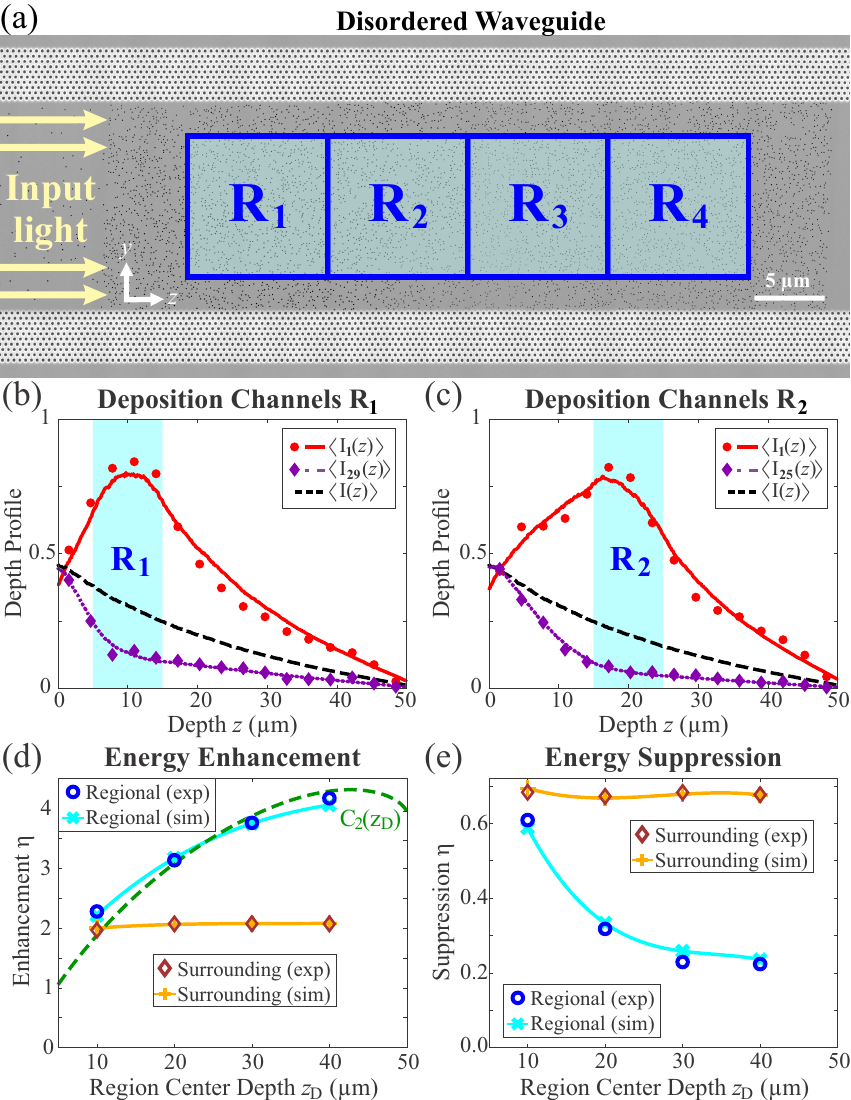}
\caption{\label{fig:experimental_measurement}
{\bf Experimental measurement of deposition eigenchannels.} (a) A composite scanning electron microscope (SEM) image of a disordered waveguide of width $W$ = 15 \textmu m. Randomly distributed air holes (each 100 nm in diameter) are etched throughout a designated $L =50$ \textmu m long region. Superimposed are four target regions used for energy deposition; each is 10 \textmu m  $\times$ 10 \textmu m.  
(b,c) Depth profiles (cross-section integrated intensities) of two deposition eigenchannels with enhanced and suppressed energies in the target region $R_1$ centered at depth $z_D$ = 10 \textmu m (b), and $R_2$ at $z_D$ = 20 \textmu m (c). Experimental data (red circle, purple diamond) agree with numerical simulations (red solid line, purple dotted line). Black dashed line is the intensity profile averaged over random input wavefronts. Each experimental data point is averaged over $\Delta z = \ell$ to reduce fluctuations. 
(d,e) Experimentally measured energy enhancement in the target region $\eta_t$ (blue-circles) and in the surrounding area $\eta_s$ (brown-diamonds) of two deposition eigenchannels $\alpha=2$ (d) and $\alpha=24$ (e) are compared with numerical data (light-blue and orange lines): for the case of energy deposition into four target regions centered at 10, 20, 30, and 40 \textmu m. In (d) the green line corresponds to $35 C_{2}(z_D)$, and its agreement with the experimental/numerical results of $\eta_t(z_D)$ confirms the essential contribution of long-range intensity correlation to energy deposition, as predicted by our analytic model.
}
\end{center}
\end{figure}

\newpage
\begin{figure}[H]
\begin{center}
\includegraphics[width=4in]{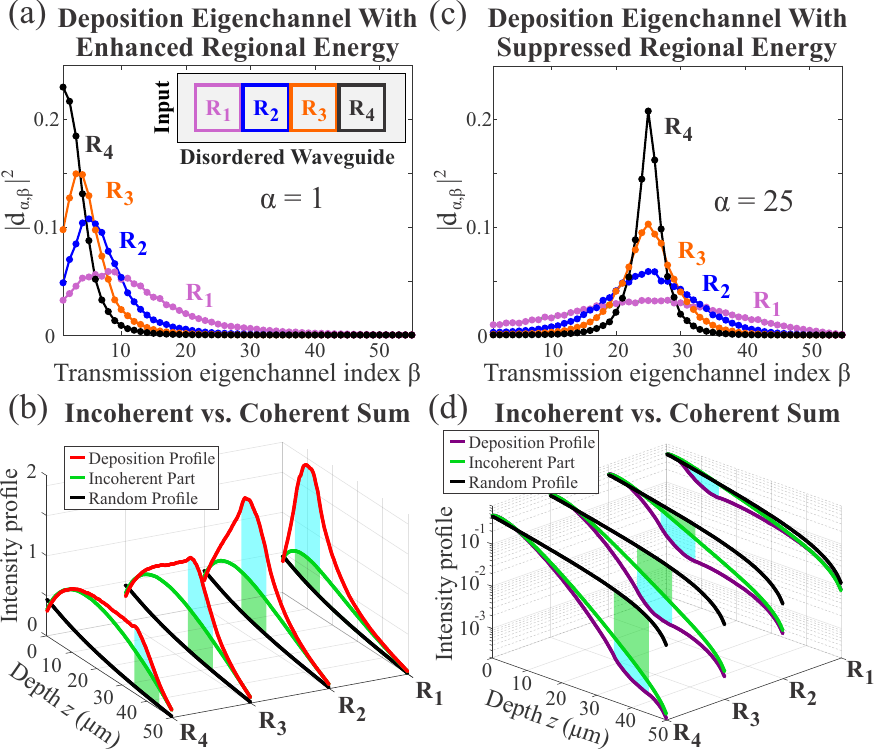}
\caption{\label{fig:composition_of_DE} 
{\bf Relation between deposition eigenchannels and transmission eigenchannels.} 
(a,c) Projection of a deposition eigenchannel with index $\alpha$ = 1 (a) or 25 (c) onto transmission eigenchannels with index $\beta$ gives the coefficients $d_{\alpha  \beta}$. Four curves denote $|d_{\alpha  \beta}|^2$ for four target regions $R_1-R_4$ [inset of (a)] in the same disordered waveguide as in Fig.~3.
(b,d) Comparison of depth profiles between coherent sum (red/purple) and incoherent (green) sum of the transmission eigenchannels with coefficients given in (a,c). While the coherent sum reproduces the deposition eigenchannel profile, the incoherent sum falls short, and their difference is attributed to interference between transmission eigenchannels. For each deposition region, enhancement/suppression above/below the random input intensity profile (black dashed line) has two distinct contributions from selective excitation of transmission eigenchannels (green areas) and constructive/destructive interference between them (cyan areas).} 
\end{center}
\end{figure}

\newpage

\section*{Supplementary materials}

Supplementary Text\\
Figs. S1 to S6\\

\renewcommand{\theequation}{S\arabic{equation}}
\setcounter{equation}{0}
\renewcommand{\thefigure}{S\arabic{figure}}
\setcounter{figure}{0}
\renewcommand{\thetable}{S\arabic{table}}
\setcounter{table}{0}

\section{Numerical simulations}

We use the Kwant simulation package~\cite{2014_Groth_Kwant} to perform numerical simulations of wave transport in a two-dimensional (2D) rectangular waveguide geometry, see Refs.~\cite{2016_Sarma_Open_Channels,2017_Koirala_Inverse_Design,2020_Bender_correlations_PRL}. The geometry of the numerical simulations is chosen to match the experimental parameters of $W/\lambda$ (width of the waveguide normalized by wavelength), $L/\lambda$ (length of the waveguide), and $N$ (number of waveguide modes). The refractive index in the input (empty) waveguide matches the average index in the disordered region, thus the number of propagating modes in the disordered waveguide is also $N$. Furthermore, the strength of the disorder and (spatially uniform) absorption coefficient are selected~\cite{2016_Sarma_Open_Channels,2020_Bender_correlations_PRL} to match the macroscopical physical parameters in the experiment: specifically the transport mean free path $\ell$ and diffusive absorption length $\xi_a$. We also simulate the disordered waveguides without loss by setting $\xi_a = \infty$. Statistical averaging over $1000$ disorder configurations is performed for all numerical results shown in Figs.~2-4 of the main text.

\subsection{Transmission eigenchannels}

We calculate the field transmission matrix $t$ in the basis of the empty (input) waveguide modes. $t$ is normalized so that when light with a unit flux in the $n$-th waveguide mode is incident on the disordered region, $\left|t_{mn}\right|^2$ is equal to the amount of flux carried away by the $m$-th waveguide mode in transmission. We also compute the wavefunction $E_n(y,z)$ describing the complex field distribution throughout the system, when excited via the $n$-th waveguide mode. 

Transmission eigenchannels are computed by performing a singular value decomposition of the transmission matrix so that $t_{mn}=\sum_{\alpha=1}^{N}U_{m\alpha}^{(T)}\cdot\tau_{\alpha}^{1/2}\cdot V_{\alpha n}^{(T)*}$. Here, $\hat{U}^{(T)}$ and $\hat{V}^{(T)}$ are unitary matrices and $\tau_{\alpha}$ are the transmission eigenvalues. The disorder-specific incident wavefront given by the $\alpha$-th column of the matrix $\hat{V}^{(T)}$ excites the $\alpha$-th transmission eigenchannel with the field distribution $E_{\alpha}^{(T)}(y,z)=\sum_{n=1}^{N}V_{\alpha n}^{(T)} \, E_{n}(y,z)$ inside the system with the transmittance given by $\tau_{\alpha}$. The depth intensity profile is computed by integrating over the transverse coordinate $y$ followed by averaging over disorder realizations: denoted by angular brackets $I_{\alpha}^{(T)}(z) = \left\langle \int_{0}^{W}\left|E_{\alpha}^{(T)}(y,z)\right|^2 \, dy\right\rangle$.

\subsection{Deposition matrix}

We provide two definitions for the deposition matrix $\mathcal{Z}$ in Eqs.~(1) and (2) of the main text. While the first definition is more general, the second one reduces to the transmission matrix at the output.
For both definitions, the deposition eigenchannels are introduced based on the singular value decomposition of the deposition matrix 
\begin{equation}
\mathcal{Z}_{mn}(z_D) =\sum_{\alpha=1}^{N}U_{m\alpha}^{(D)}(z_D) \, \zeta_{\alpha}^{1/2}(z_D)\, V_{\alpha n}^{(D)*}(z_D).
\end{equation} 
The spatial structure of $\alpha$-th deposition eigenchannel inside the system is given by $E_{\alpha}^{(D)}(y,z;z_D)=\sum_{n=1}^{N}V_{\alpha n}^{(D)}(z_D)\, E_{n}(y,z)$. The depth intensity profile is computed by integrating over the transverse coordinate $y$ as well as the disorder realizations $I_{\alpha}^{(D)}(z;z_D)= \left\langle \int_{0}^{W}\left|E_{\alpha}^{(D)}(y,z;z_D)\right|^2 dy \right\rangle$.

\begin{figure}[H]
\begin{center}
\includegraphics[width=1\linewidth]{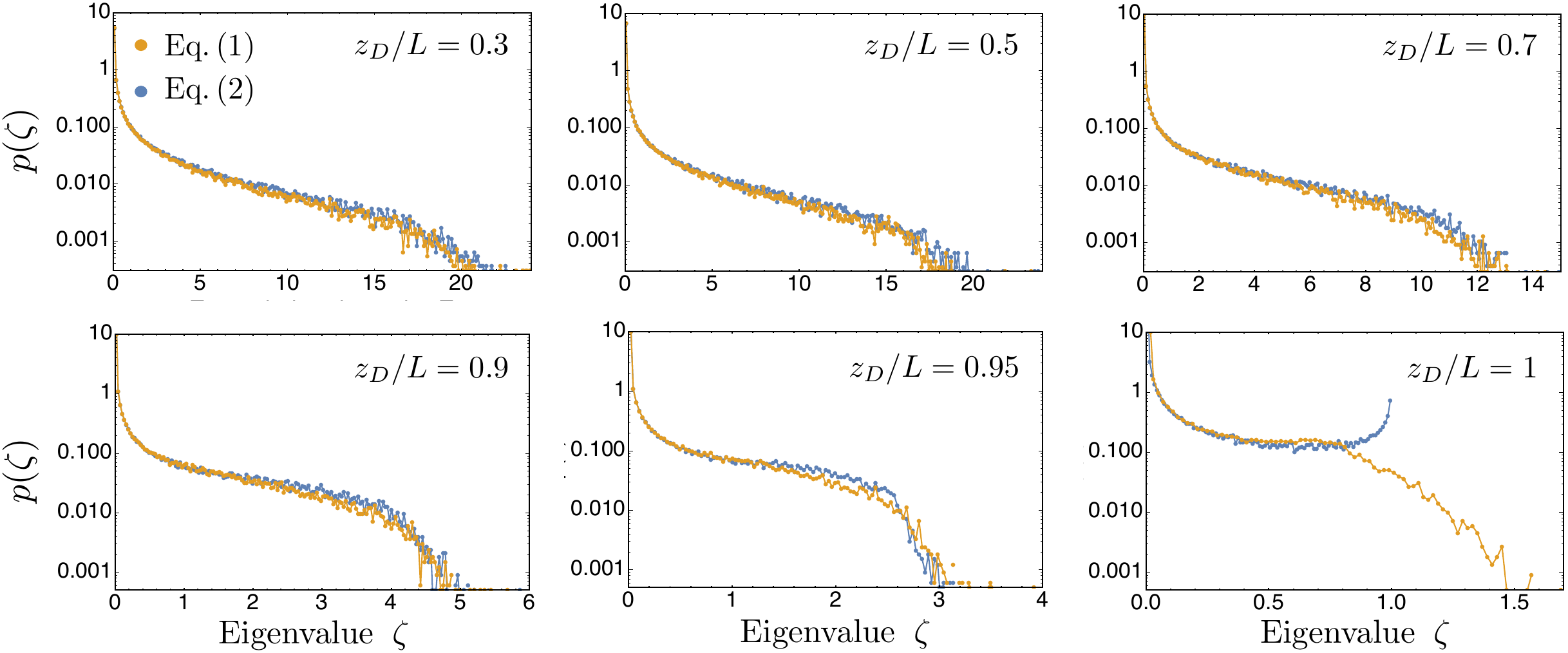}
\caption{
\label{FIGS6} 
Comparison of the eigenvalue distributions $p(\zeta)$ of the operators $\mathcal{Z}^{\dagger}\mathcal{Z}$ defined by Eq.~(\ref{eq:Z_definition1}) and Eq.~(\ref{eq:Z_definition2}), evaluated at different depths $z_D/L$ of a disordered waveguide (length $L=50\, \mu$m,  width $W= 30\, \mu$m, transport mean free path $\ell=3.3 \, \mu$m). Noticeable differences are observed only at $z_D$ very close to $L$, where $p(\zeta)$ converges to the bimodal distribution of transmission eigenvalues for the operator $\mathcal{Z}^{\dagger}\mathcal{Z}$ defined by Eq.~(\ref{eq:Z_definition2}) only. The two distributions still coincide for $z_D/L=0.95$ (panel 5), which corresponds to $L-z_D<\ell$.
} 
\end{center}
\end{figure}

Numerically we compare the eigenvalues $\zeta(z_D)$ of the deposition matrices $\mathcal{Z}(z_D)$ defined by Eqs.~(1) and (2) for a thin slice at depth $z_D$ inside the disordered waveguide. As shown in Fig.~\ref{FIGS6}, the probability density function (PDF) of deposition eigenvalues $P(\zeta)$ is almost identical for the two definitions at most depths inside the disordered waveguide. Only close to the very end $L-z_D < \ell$ do the two PDFs differ; one remains single peaked while the other becomes bimodal and converges to the PDF of the transmission eigenvalues.   

To illustrate the close relationship between the two definitions of the deposition matrix, we compare the trace of $\mathcal{Z}^{\dagger}\mathcal{Z}$, which corresponds to the sum of their eigenvalues, $ Tr[\mathcal{Z}^{\dagger}\mathcal{Z}] =\sum_m \zeta_m$. For the first definition, we switch to the waveguide mode basis and find the trace $Tr[\mathcal{Z}(z_D)^{\dagger}\mathcal{Z}(z_D)] = \sum_{n=1}^N\sum_{m=1}^{\infty} \left| \mathcal{Z}_{mn}(z_D) \right|^2$, where $\mathcal{Z}_{mn}(z_D)=\int_{0}^{W}\chi_m(y) \, E_n(y,z_D) \, dy$ is obtained by projecting the internal field distribution, excited by a unit flux input to the $n$-th waveguide mode, onto the $m$-th waveguide mode at the cross-section $z=z_D$. With the second definition of $\mathcal{Z}(z_D)$, the trace $Tr[\mathcal{Z}^{\dagger}(z_D)\mathcal{Z}(z_D)] = \sum_{n=1}^N\sum_{m=1}^{N} v_m \, \left|\mathcal{Z}_{mn}(z_D)\right|^2$ differs from the first one in two ways: (i) the summation over $m$ runs only over the propagating modes of the waveguide, and (ii) the prefactor $v_m$ introduces a weight for different modes. Using the Fisher-Lee formula~\cite{1981_Fisher}, one can show that the trace for the second definition at $z_D=L$ is equal to the dimensionless conductance $g=\sum_m\tau_m$.

\section{Analytical predictions}

\subsection{Filtered random matrix (FRM) model}

\begin{figure}[H]
\begin{center}
\includegraphics[width=1\linewidth]{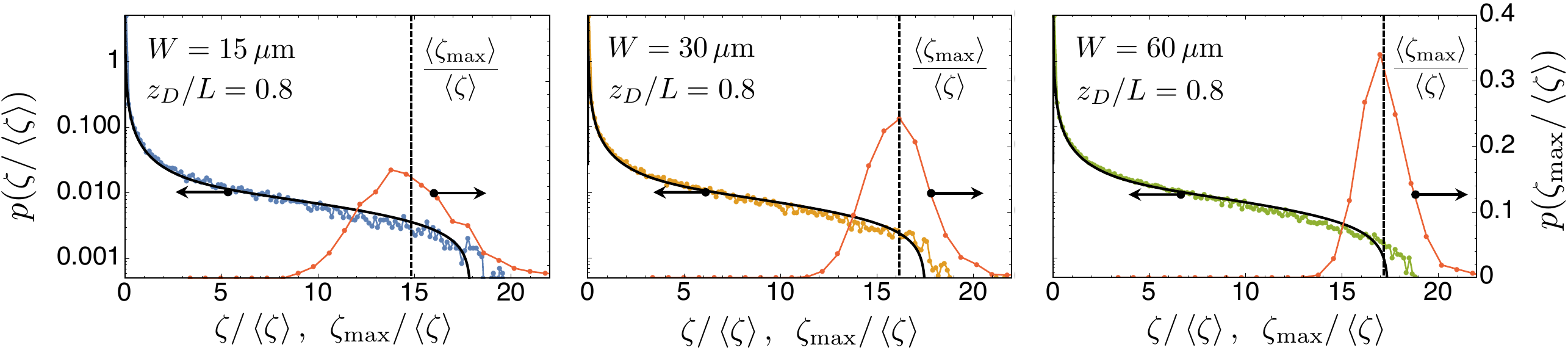}
\caption{
\label{FIGS3} 
Deposition eigenvalue distribution $p(\zeta)$ at depth $z_D=0.8L$ of a disordered waveguide of length $L=50\, \mu$m and  width $W=15, 30, 50\, \mu$m. Analytical FRM predictions (solid black lines) are compared with numerical results (dots) obtained from the solution of the wave equation for $10^3$ realizations of the disordered waveguide with a transport mean free path $\ell=3.3 \, \mu$m. The distribution of the largest eigenvalue $p(\zeta_\text{\rm max})$ is superimposed (red dots connected by red line) to reveal the convergence of $\zeta_\text{\rm max}$ towards the upper edge of $p(\zeta)$ in the limit $g=N\left<\tau \right> \gg1$ ($W=15, 30, 50\, \mu$m correspond to $g=5, 10, 15$). The value  $\left<\zeta_\text{\rm max} \right>$ is indicated with dashed vertical line.} 
\end{center}
\end{figure}

In the main text, we make the ansatz that $\mathcal{Z}(z_D)$ has the same spectrum as a filtered matrix $\tilde{t}$ drawn from a larger virtual transmission matrix $t_0$. The matrix $\tilde{t}$ is obtained by keeping a fraction $m<1$ of rows and columns in $t_0$.  In Ref.~\cite{2013_Goetschy}, it is shown that the eigenvalue distribution of the matrix $\tilde{t}^\dagger \tilde{t}$ is given by $p_{\tilde{t}^\dagger \tilde{t}}(x)= - \lim_{\eta \to 0^+} \mathrm{Im} g_{\tilde{t}^\dagger \tilde{t}}(x + \textrm{i} \eta)$, where the resolvent $g_{\tilde{t}^\dagger \tilde{t}}(w)$ is solution of the implicit equation:
\begin{equation}
\label{FRM}
g_{\tilde{t}^\dagger\tilde{t}}(w)=
\frac{w\,m\,g_{\tilde{t}^\dagger\tilde{t}}(w)+1-m}{w\,m^2\,g_{\tilde{t}^\dagger\tilde{t}}(w)}g_{t_0^\dagger t_0}
\left[
\frac{\left[w\,m\,g_{\tilde{t}^\dagger\tilde{t}}(w)+1-m\right]^2}{w\, m^2\,g_{\tilde{t}^\dagger\tilde{t}}(w)^2}
\right].
\end{equation}
Since $t_0$ represents the transmission matrix of a virtual opaque and disordered medium, the resolvent $g_{t_0^\dagger t_0}(w)$ is~\cite{2013_Goetschy}:
 \begin{equation}
 \label{ResolventTM}
g_{t_0^\dagger t_0}(w)=\frac{1}{w}-\frac{\bar{\tau}_0}{w\sqrt{1-w}}\textrm{Arctanh}\left[
\frac{\textrm{Tanh}(1/\bar{\tau}_0)}{\sqrt{1-w}}\right],
\end{equation}
where $\bar{\tau}_0$ is the mean of $p_{t_0^\dagger t_0}(x)$. Hence, the eigenvalue distribution $p(\zeta)$ of $\mathcal{Z}^\dagger (z_D)\mathcal{Z}(z_D)$ is parametrized by $m$ and $\bar{\tau}_0$ only. In particular, the variance of $p(\zeta)$ is
\begin{equation}
\frac{\text{Var}(\zeta)}{\left<\zeta\right>^2}=m\left(\frac{2}{3\bar{\tau}_0} -1 \right)+1-m.
\label{EqVariance}
\end{equation} 
In our model, we take $m=\left<\zeta(L)\right>/\left<\zeta(z_D)\right> = \left<\tau \right>/\left<\zeta(z_D)\right> \le 1$ and $\bar{\tau}_0$ solution of Eq.~\eqref{EqVariance}. In this way, the full distribution $p(\zeta)$ becomes parametrized by its first two moments, $\left<\zeta \right>$ and $\left<\zeta^2 \right>$. To obtain the FRM predictions in Fig.2(a) of the main text and Fig.~\ref{FIGS3} here, we solve Eq.~\eqref{FRM} with $\left<\zeta \right>$ and $\left<\zeta^2 \right>$ as input parameters evaluated numerically. The good agreement between the FRM prediction and numerical distributions validates our model.  

\begin{figure}[H]
\begin{center}
\includegraphics[width=0.7\linewidth]{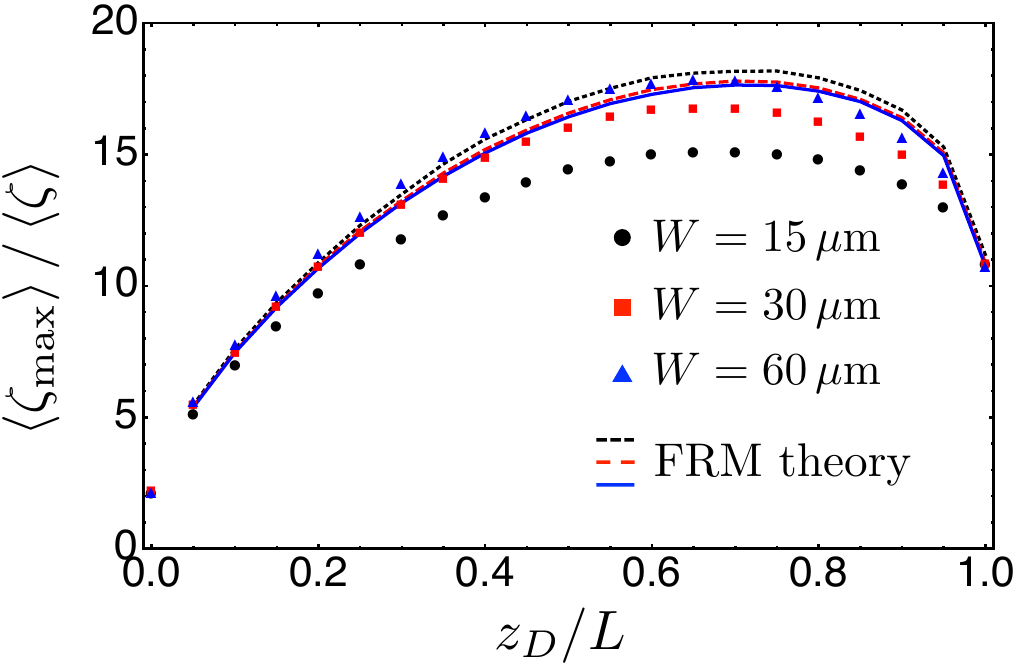}
\caption{
\label{FIGS4} 
Symbols (circles, squares, triangles) represent the ensemble average of the largest deposition eigenvalue $\left<\zeta_\text{\rm max} \right>$ at different depth $z_D/L$ for three waveguide width $W$ = 10, 30, 60 $\mu$m. Other parameters are identical to the paremeters in Fig.~\ref{FIGS3}. Solid lines of matched colors are analytical predictions for the upper edge of $p(\zeta)$ evaluated with the numerical mean $\langle \zeta \rangle$ and variance $\text{Var}[\zeta]$. The agreement between the numerical data and the analytical predictions improves with increasing waveguide width. 
} 
\end{center}
\end{figure}

In Ref.~\cite{2013_Goetschy}, it is also shown that the edges $x^*$ of the distribution $p_{\tilde{t}^\dagger \tilde{t}}(x)$ are given by $x^*=\xi^*\left[1+(m-1)/\xi^*g_{t_0^\dagger t_0}(\xi^*)\right]^2$, where $\xi^*$ is the solution of
\begin{equation}
\label{SolEdge}
\left.\frac{\textrm{d}g_{t_0^\dagger t_0}(\xi)}{\textrm{d}\xi}\right\vert_{\xi^*}=\frac{g_{t_0^\dagger t_0}(\xi^*)}{2\xi^*} \,
\frac{-(1-m)^2+\xi^{*2}g_{t_0^\dagger t_0}(\xi^*)^2}{(1-m)^2-(1-m)\xi^*g_{t_0^\dagger t_0}(\xi^*)}.
\end{equation}
We solve this equation to find the values of the upper edge $x^*$ represented in Fig.~2(b) of the main text, where it is compared to $\left<\zeta_\text{\rm max} \right>$. In the limit of large matrix size ($N\to \infty$), we expect that the upper edge of $p(\zeta)$ and $\left<\zeta_\text{\rm max} \right>$ coincide. This is illustrated in Fig.~\ref{FIGS3}, where we present the distributions $p(\zeta)$ and $p(\zeta_\text{\rm max})$ for three waveguide widths $W$, at a fixed depth $z_D=0.8L$. As $W$ increases, $p(\zeta)$ is almost unaffected because $\text{Var}(\zeta)$ marginally depends on $W$, whereas the distribution $p(\zeta_\text{\rm max})$ shrinks and $\left<\zeta_\text{\rm max} \right>$ converges towards the upper edge from below. Convergence is reached for all depths $z_D$ in the limit of large conductance ($g=N\left<\tau \right>\gg 1$), as illustrated in Fig.~\ref{FIGS4}.

\subsection{First two moments of $p(\zeta)$}

\begin{figure}[H]
\begin{center}
\includegraphics[width=1\linewidth]{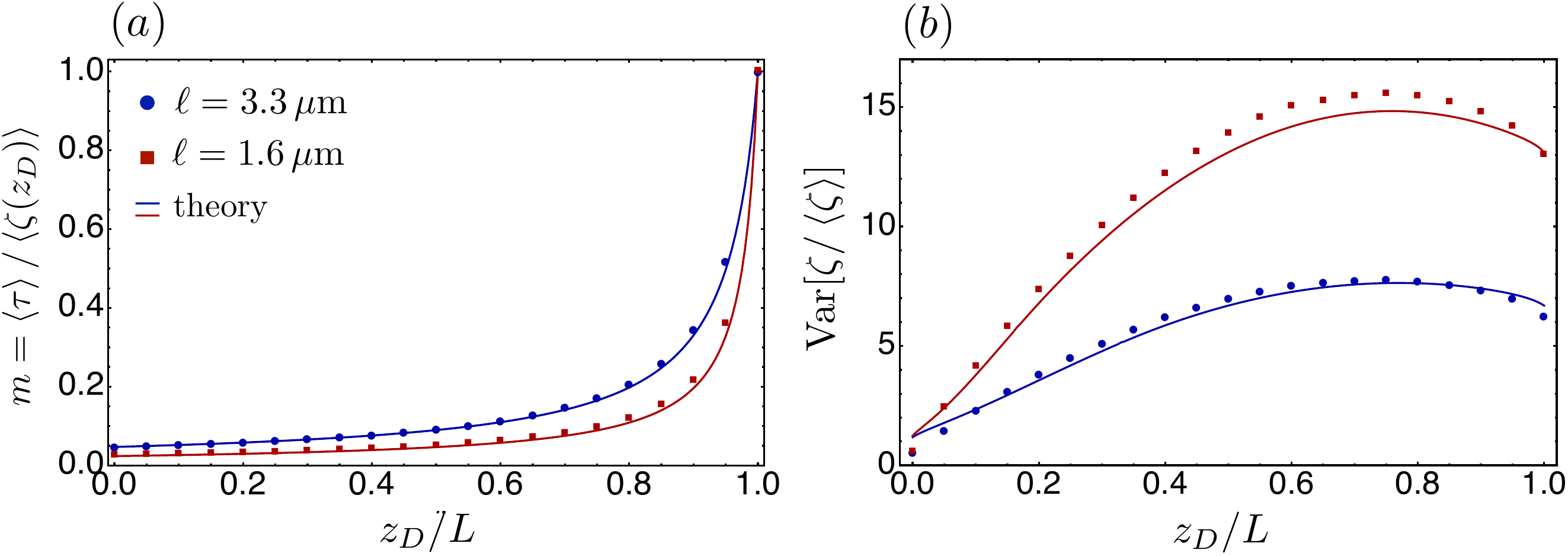}
\caption{
\label{FIGS5} 
(a) Effective filtering ratio $m=\left<\tau\right>/\left< \zeta (z_D) \right>$ of the FRM model versus depth $z_D/L$. Numerical results (dots) are compared with the analytical prediction $m=1/\left[2(1/\left<\tau\right>-1)(1-z_D/L)+1\right]$ (solid lines with matched colors); (b) Variance $\textrm{var}[\zeta/\langle \zeta \rangle] = \left<\zeta^2\right>/\left<\zeta\right>^2-1$ of the eigenvalue distribution $p(\zeta)$ vs. depth $z_D/L$. Numerical results (dots) are compared with intensity fluctuations $NC_2(z_D) +1$ evaluated analytically (solid lines of matched colors) for two values of transport mean free path $\ell$ = 1.6, 3.3 $\mu$m. The disordered waveguide dimensions are $L= 50\, \mu$m and $W= 15\, \mu$m.
} 
\end{center}
\end{figure}

The first moment $\left<\zeta(z_D) \right>$ of the distribution $p(\zeta)$ is proportional to the mean intensity $\left<I(z_D) \right>$ deposited at depth $z_D$ under random wavefront illumination. We can approximate it by the steady state solution of the diffusion equation with an isotropic source located at an injection depth $z_\text{in}\sim \ell$ away from the front surface of the disordered waveguide boundary, $\partial^2_z\left<\zeta(z) \right>=A\delta(z-z_\text{in})$, where $A$ is a constant to be evaluated below. This equation must be complemented with boundary conditions: $\left<\zeta(z=0) \right>=z_0\partial_z\left<\zeta(z=0) \right>$  and $\left<\zeta(z=L) \right>=-z_0\partial_z\left<\zeta(z=L) \right>$, where $z_0$ is the extrapolation length ($z_0=\pi \ell/4$ in 2D and $z_0=2 \ell/3$ in 3D). The solution is a linear function of $z_D$,
\begin{equation}
\left<\zeta(z_D) \right>=A\frac{(z_\text{in}+z_0)(L+z_0-z_D)}{L+2z_0}, 
\end{equation}
for $z_D > z_\text{in}$. The constant $A$ being fixed by our choice of normalization $\left<\zeta(z=L) \right>=\left<\tau \right>=2z_0/(L+2z_0)$, we get
\begin{equation}
\left<\zeta(z_D) \right>=2(1-\left<\tau\right>)\left(1-\frac{z_D}{L}\right)+\left< \tau \right>,
\end{equation}
which is independent of the precise value of $z_\text{in}$. The agreement of this prediction with numerical simulations is excellent, as shown in Fig.~\ref{FIGS5}(a).  

The variance $\text{Var}(\zeta)=\left<\zeta(z_D)^2 \right>-\left<\zeta(z_D) \right>^2$ of the eigenvalue distribution $p(\zeta)$ can be related to intensity fluctuation $\left<I(z_D)^2\right>-\left<I(z_D)\right>^2$. Using the singular value decomposition of the deposition matrix $\mathcal{Z}=U^{(D)}\hat{\zeta}^{1/2}V^{(D) \, \dagger}$, the cross-section integrated intensity deposited by a waveguide mode $n$ is $I_n(z_D)=\sum_\alpha\vert V^{(D)}_{n \alpha}\vert^2\zeta_\alpha(z_D)$. The evaluation of the first two moments of $I_n(z_D)$ is straightforward using the isotropy hypothesis for the disordered waveguide~\cite{1990_Mello, 1997_Beenakker}. This amounts to considering that $V$ is uniformly distributed over the unitary group and is independent of $\hat{\zeta}$. We find
\begin{align}
\left<I_n(z_D)\right>&=\frac{1}{N}\left<\text{Tr}(\hat{\zeta})\right>,
\\
\left<I_n^2(z_D)\right>&=\frac{1}{N^2-1}\left(1-\frac{1}{N}\right)\left[\left<\text{Tr}(\hat{\zeta})^2\right>+ \left<\text{Tr}(\hat{\zeta}^2)\right>\right].
\end{align}
In the limit $N\gg1$, the leading order is 
\begin{align}
\text{Var}[I_n(z_D)]&\simeq\frac{1}{N^2}\left[ \left<\text{Tr}(\hat{\zeta}^2)\right>-\frac{1}{N}\left<\text{Tr}(\hat{\zeta})^2\right> \right]
\nonumber
\\
&\simeq \frac{1}{N^2}\left[ \left<\text{Tr}(\hat{\zeta}^2)\right>-\frac{1}{N}\left<\text{Tr}(\hat{\zeta})\right>^2 \right].
\end{align}
 This result is independent of the waveguide mode index $n$, and also holds for random wavefront illumination. We conclude that
\begin{equation}
\frac{ \text{Var}[\zeta(z_D)]}{\left<\zeta(z_D) \right>^2}\simeq N\frac{ \text{Var}[I(z_D)]}{\left<I(z_D) \right>^2}.
\label{EqVar}
\end{equation}

Finally, the intensity fluctuations at depth $z_D$ are computed by decomposing the field $E(z_D)$ involved in $I(z_D)=\vert E(z_D) \vert^2$ as a sum of propagators along all possible scattering trajectories~\cite{1999_VanRossum}. The intensity fluctuations are composed of a small Gaussian field contribution $C_1=1/N$, and dominated by the non-Gaussian contribution $C_2(z_D)$,
\begin{equation}
\frac{ \text{Var}[I(z_D)]}{\left<I(z_D) \right>^2}=C_1+C_2(z_D),
\label{EqVar2}
\end{equation} 
with 
\begin{equation}
C_2(z)=\frac{2}{gL\left<I(z) \right>^2}\int_0^L dz' \left<I(z') \right>^2\left[\partial_{z'} K(z,z') \right]^2.
\label{EqC2}
\end{equation}
The mean intensity is $\left<I(z) \right>=\int_0^L dz'e^{-z'/\ell}K(z,z')$, where $K(z,z')$ is the Green's function of the diffusion equation $\partial^2_zK(z,z')=\delta(z-z')$, with boundary conditions
$\partial_z K(0,z')= K(0,z')/z_0$ and $\partial_z K(L,z')= -K(L,z')/z_0$. The solution is
\begin{equation}
K(z,z')=\frac{(z^-+z_0)(L+z_0-z^+)}{L+2z_0},
\end{equation}
with $z^-=\text{min}(z,z')$ and $z^+=\text{\rm max}(z,z')$. In the limit $L \gg \ell$, the 
correlator $C_2(z)$ takes the simple form~\cite{2014_Sarma}
\begin{equation}
C_2(z)\simeq \frac{2}{3g}\frac{z(3L-2z)}{L^2},
\end{equation} 
where $g=N\left<\tau\right>$ is the dimensionless conductance of the disordered waveguide. 

By combining Eqs.~\eqref{EqVar} and~\eqref{EqVar2}, we finally obtain an analytical expression for the normalized variance of the eigenvalues of the deposition matrix,
\begin{equation} 
\frac{ \text{Var}[\zeta(z_D)]}{\left<\zeta(z_D) \right>^2}\simeq 1 +NC_2(z_D).
\end{equation}
Figure~\ref{FIGS5}(b) shows a good agreement between the simulation results and our prediction based on Eq.~\eqref{EqC2}.

When comparing with the experimental data in Fig.~\ref{fig:experimental_measurement}(d), the effect of absorption is included in Eq.~(\ref{EqC2}). This is accomplished by substitution of the Green's function which accounts for absorption
\begin{equation}
\partial_z^2 K(z,z^\prime) -\frac{K(z,z^\prime)}{\xi_a^2}=\delta(z-z^\prime),
\label{eq:kernel_equation}
\end{equation}
where $\xi_a$ is the diffusive absorption length. 

\subsection{Upper edge of $p(\zeta)$}

In the main text, we argue that the maximal enhancement of energy deposition $\left<\zeta_\text{\rm max} \right>/\left<\zeta \right>$ depends, for most depths $z_D$, only on the long-range intensity-intensity correlation function $C_2(z_D)$. To prove this property, we first note that $m=\left<\tau \right> /\left<\zeta(z_D) \right>$ becomes quickly smaller than unity for $z_D< L$, as long as $\left<\tau \right>\ll 1$ [see Fig.~\ref{FIGS5}(a)]. This allows us to perform an expansion of the FRM solution in the limit $m\to 0$. Using 
\begin{equation}
g_{t_0^\dagger t_0}(w)\simeq \frac{1-\bar{\tau}_0}{w}-\frac{i\pi\bar{\tau}_0}{2w\sqrt{1-w}},
\end{equation}
and expanding Eq.~\eqref{SolEdge} to leading order, we find
\begin{equation}
\label{SolEdge2}
\frac{\left<\zeta_\text{\rm max}(z_D) \right>}{\left<\zeta(z_D)\right>}\simeq \frac{[(\gamma-1)^{2/3}+(\pi/2)^{2/3}]^2[\gamma-1+(\pi/2)^{2/3}(\gamma-1)^{1/3}]}{\gamma(\gamma-1)^{1/3}}
+ \mathcal{O}(m),
\end{equation}
which depends on $\gamma=m/\bar{\tau}_0$ only. According to Eq.~\eqref{EqVariance}, 
\begin{equation}
\gamma \simeq \frac{3}{2} \left(\frac{\text{Var}[\zeta(z_D)]}{\left<\zeta (z_D)\right>^2} -1\right)
\simeq \frac{3NC_2(z_D)}{2}.
\end{equation}
 Hence,  $\left<\zeta_\text{\rm max} \right>/\left<\zeta \right>$ depends only on $C_2(z_D)$ only. Since $\gamma \sim NC_2(z_D) \gg 1$ for $\left<\tau\right>\sim \ell/L \ll 1$, we can further expand Eq.~\eqref{SolEdge2} as
 \begin{equation}
\frac{\left<\zeta_\text{\rm max}(z_D) \right>}{\left<\zeta(z_D) \right>} \simeq \gamma + 3\left(\frac \pi 2\right)^{2/3}\gamma^{1/3}-2 + \mathcal{O}(\gamma^{-1/3}).
\end{equation}
This shows that the energy enhancement slowly converges to $\gamma$ in the limit $L/ \ell \gg 1$.

\section{Experimental Measurements}

\subsection{Sample fabrication}

We fabricate the two-dimensional (2D) waveguide structures on a silicon-on-insulator wafer with electron-beam lithography and reactive ion etching. The entire structure includes an ridge-waveguide and a tapered waveguide for light injection, a buffer region and the primary disordered section.  The spatially modulated light is injected from the side/edge of the wafer into the ridge waveguide (width = 300 \textmu m, length = 15 mm). It then enters the tapered waveguide (tapering angle = $15^{\circ}$). The tapered waveguide width decreases gradually from 300 \textmu m to 15 \textmu m. The tapering results in waveguide mode coupling and conversion~\cite{2016_Sarma_Open_Channels}. To avoid light leakage, from the tapered waveguide onward the sidewalls are photonic crystals, with a complete 2D bandgap, to confine light in the waveguides. Specifically, the photonic-crystal boundaries consist of 16-layered triangle-lattices of air holes (radius = 155 nm, lattice constant = 440 nm). They provide a 2D complete bandgap for TE polarized light (used in the experiment) within the wavelength range of 1120 nm to 1580 nm~\cite{2014_Yamilov_PRL}. The tapered waveguide is followed by a weakly-scattering region (25 \textmu m long), denoted the ‘buffer’, followed by the multiple-scattering region (50 \textmu m long). The out-of-plane scattering from the buffer region provides information about the light incident upon the diffusive region without altering the overall diffusive transport of light in our system~\cite{2020_Bender_correlations_PRL}. Randomly distributed air holes (diameter = 100 nm) with a minimum (edge-to-edge) distance of 50 nm, are fabricated in the silicon top-layer to induce light scattering in the buffer and diffusive regions. The diffusive region has 5250 holes, which results in an air filling fraction in the Si of $5.5 \% $. The number of air holes in the buffer region is 260, and the air filling fraction is $0.55 \% $.

\subsection{Optical setup}

\begin{figure}[H]
\begin{center}
\includegraphics[width=5in]{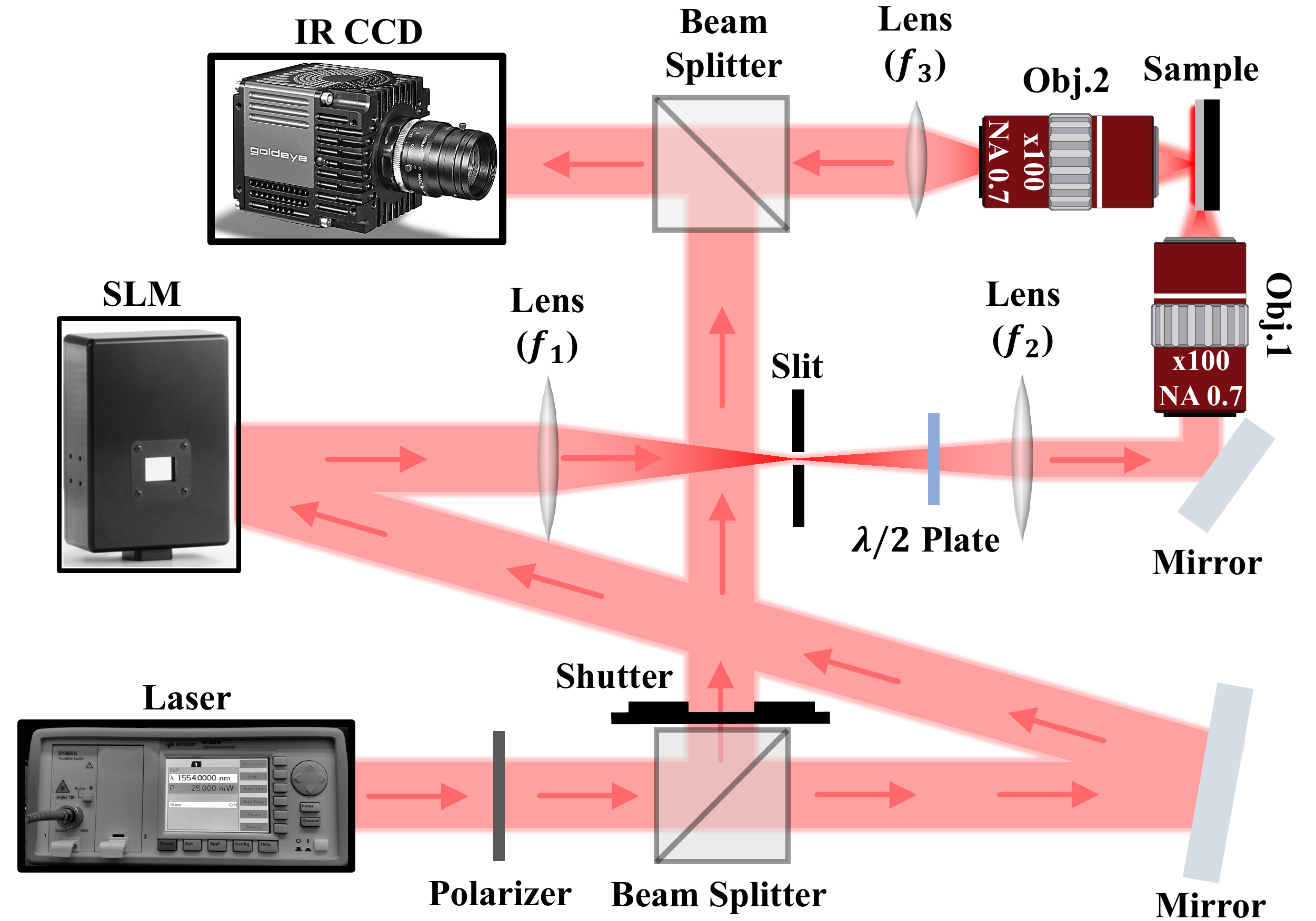}
\caption{\label{fig:setup} 
{\bf A detailed schematic of our experimental setup} 
Monochromatic light from our CW laser is linearly polarized and split into two beams. One beam illuminates the phase modulating region of a spatial light modulator (SLM). The other beam is used as a reference. The SLM controls the input wavefront injected into our diffusive waveguides. A beam splitter merges the light collected from the top of our sample with the reference beam and their interference pattern is recorded by an IR CCD. The focal length of the three lenses used in this setup are: $f_{1}=400$ mm, $f_{2}=75$ mm, and $f_{3}=100$ mm.} 
\end{center}
\end{figure}

A detailed schematic of our experimental setup is presented in Fig.~\ref{fig:setup}. A wavelength tunable laser (Keysight 81960A) outputs a linearly-polarized continuous-wave (CW) beam with a wavelength around 1554 nm. The collimated beam is split with a 50/50 beam splitter into two beams. One is used as a reference beam, while the other illuminates the phase modulating surface of a phase-only SLM (Hamamatsu LCoS X10468). Displayed on the SLM is an one-dimensional (1D) phase-modulation pattern consisting of 128 macropixels. Each micropixel has $4\times800$ SLM pixels. We image the field reflected from the SLM plane onto the back focal plane of a long-working-distance objective Obj. 1 (Mitutoyo M Plan APO NIR HR100$\times$, Numerical Aperture = 0.7) using two lenses with focal lengths of $f_{1}=400$ mm and $f_{2}=75$ mm. To prevent the unmodulated light from entering the objective lens, we display a binary diffraction grating within each macropixel to shift the modulated light away from the unmodulated light in the focal plane of the first lens $f_{1}$. With a slit in the focal plane, we block all light except the phase-modulated light in the first diffraction order. Right after the slit and before the second lens $f_{2}$, we insert a half-wave $(\lambda/2)$ plate to rotate the polarization of light so that it is transverse-electric (TE) polarized relative to our waveguide sample. The waveguide entrance at the edge of our SOI wafer is placed at the front focal plane of Obj. 1, so that it is illuminated with the Fourier transform of the phase-modulation pattern displayed on the SLM. From the top of the wafer, another long-working-distance objective (Obj. 2 (Mitutoyo M Plan APO NIR HR100$\times$) collects light scattered out-of-plane from the waveguide. We use a third lens with a focal length of $f_{3}=100$ mm together with Obj. 2 to image the sample. In conjunction, the lens and the objective magnify the sample image by $50$ times. Using a second beam splitter, we combine the light collected from the sample with the reference beam. Their interference patterns are recorded with an IR CCD camera (Allied Vision Goldeye G-032 Cool).

\subsection{Deposition matrix measurement}

With the interferometric setup described in the last subsection, we can determine the field distribution of light within the disordered region of the waveguide from the out-of-plane scattered light: for any phase-modulation pattern displayed on the SLM. To do this, for a given phase modulation pattern on the SLM, we first measure the 2D intensity distribution of light inside the waveguide by blocking the reference beam with a shutter. Then after unblocking the reference beam, we retrieve the phase profile of the scattered light with a four-phase measurement. In this measurement, the global phase of the pattern displayed on the SLM is modulated four times in increments of $\pi/2$ rad ~\cite{CCCL}. 

For a given disordered waveguide configuration, we sequentially apply a complete set of orthogonal phase-fronts to the SLM and record the 2D field distribution throughout the waveguide for each input. Based on these field measurements, we construct a linear matrix that relates the field pattern at the SLM to the field distribution anywhere inside the waveguide. In particular, we create two matrices $\mathcal{Z}_{\rm SLM\rightarrow Buff}$ and $\mathcal{Z}_{\rm SLM\rightarrow R}$, which map the field from the SLM surface to the buffer region in front of the main disordered region and to the deposition area of interest $R$ inside it, respectively.  

With these two matrices, we build the deposition matrix relating the field in the buffer region to the field in the deposition region: $\mathcal{Z}_{\rm Buff\rightarrow R}\equiv \mathcal{Z}_{\rm SLM\rightarrow R} \, \mathcal{Z}_{\rm SLM\rightarrow Buff}^{-1}$. To calculate the inverse of $\mathcal{Z}_{\rm SLM\rightarrow Buff}$, we use Moore-Penrose matrix inversion. The inverse matrix is calculated using the the $55$ highest singular values of $\mathcal{Z}_{\rm SLM\rightarrow Buff}$ and the remaining singular values set to zero. This restriction is imposed because our diffusive waveguide only has 55 transmission eigenchannels, and the remaining singular values are dominated by  experimental noise.

As shown in Fig.~3(a) of the main text, the energy deposition regions are four 10 \textmu m $\times$ 10 \textmu m areas inside the disordered waveguide. To avoid artifacts from light scattered out-of-plane from the photonic-crystal sidewalls, the deposition regions are kept away from the waveguide boundaries. Since each deposition area is relatively large and contains many speckle grains, the effect of incident wavefront shaping on increasing/decreasing out-of-plane scattered light into the camera is reduced. Therefore, optimizing the input wavefront to the waveguide predominantly enhance/suppress the amount of energy deposited into the target region. This is confirmed by our numerical simulation with realistic parameters, as detailed in the next subsection.  

\subsection{Deposition eigenchannel characterization}

To experimentally excite a single deposition eigenchannel, we first calculate the singular vectors of the deposition matrix, $\mathcal{Z}_{\rm Buff \rightarrow R}$, using a singular value decomposition. We then convert the singular vectors of $\mathcal{Z}_{\rm Buff \rightarrow R}$ into SLM phase-modulation patterns by multiplying each vector by the pseudoinverse of the matrix from the SLM to the buffer region, $\mathcal{Z}^{-1}_{\rm SLM \rightarrow Buff}$, and retain the resulting phase-modulation patterns. By displaying one of the phase patterns on the SLM, we excite the corresponding deposition eigenchannel in the diffusive waveguide. For a given disorder configuration and the region of interest, we record the 2D spatial intensity profiles of every deposition eigenchannel of $\mathcal{Z}_{\rm Buff \rightarrow R}$. From each eigenchannel measurement, we integrate the 2D intensity pattern over the waveguide cross-section along $y$ to obtain the deposition eigenchannel’s depth profile. While our waveguide structure has a width of 15 \textmu m, we only use the central 10 \textmu m region of the waveguide’s out-of-plane-scattered light to avoid artifacts from out-of-plane scattering by the photonic-crystal boundaries.

After measuring 55 of the deposition eigenchannel profiles of $\mathcal{Z}_{\rm Buff \rightarrow R}$ with the highest eigenvalues, for a given disorder configuration, we need to mitigate the influence of ‘noisy eigenchannels’ and properly normalize the eigenchannel profiles. While our waveguides have 55 deposition eigenchannels, for a given region of interest, the limited dynamic range of our CCD camera makes the deposition eigenchannels with small eigenvalues experimentally inaccessible. The missing information needed to reconstruct these deposition eigenchannels is replaced with measurement noise, and therefore the corresponding ‘noisy eigenchannels’ are equivalent to random inputs. Additionally, this effect leads to a slight shuffling in the order of the measured eigenchannels based on their eigenvalues. 

To account for redundant eigenchannel measurements, induced by measurement noise, as well as the inability to experimentally control the norm of the input flux, we conduct numerical simulations in order to normalize and sort the measured deposition eigenchannel depth profiles. The simulated waveguides have identical parameters and dimensions to the experimental ones. We compute the deposition matrices that map the incident fields to the fields within four target regions $R_1 - R_4$. Since $2N$ points are chosen randomly inside each region, the deposition matrix is rectangular with dimensions $2N\times N$. We compute the ensemble-averaged eigenchannel profiles from the numerical simulations to normalize and determine the correct order of measured eigenchannel profiles. First, we normalize all of the experimental and numerical profiles to have a mean value of one, and spatially overlap them. Then, for each experimental profile, we calculate its absolute difference from every numerical profile and assign the correct order based on the minimum difference. Once we have the correct order of the experimental eigenchannel, we renormalize it according to the unit input flux. In this process, we remove the ‘noisy eigenchannels’ with the intermediate eigenvalues, whose depth profiles resemble those of random input wavefronts. In this way, we are able to sort out the deposition eigenchannels of a single realization. 

We repeat this process for multiple disorder realizations -generated using different wavelengths and random hole configurations- and ensemble-average the depth profiles of deposition eigenchannels with same indices. In total, we measure the deposition eigenchannel profiles of 13 independent system realizations: for each of four target regions at different spatial locations in the waveguide. We obtain these measurements from two waveguides with different random arrays of air holes. To generate independent system realizations from the same hole configuration, we vary the wavelength of the input light beyond the spectral correlation width of the diffusive light, which is estimated to be 0.4 nm. Over a wavelength span of 3 nm, we vary the input wavelength of our laser in increments of 0.5 nm. We choose the specific wavelength range of the measurement -for each random hole configuration- such that the effective dissipation by out-of-plane scattering is minimal and nearly constant over the probe wavelength range. Because of minor structural differences, pertaining to the fabrication process, one sample has a smaller acceptable wavelength range.

\begin{figure}[H]
\begin{center}
\includegraphics[width=4in]{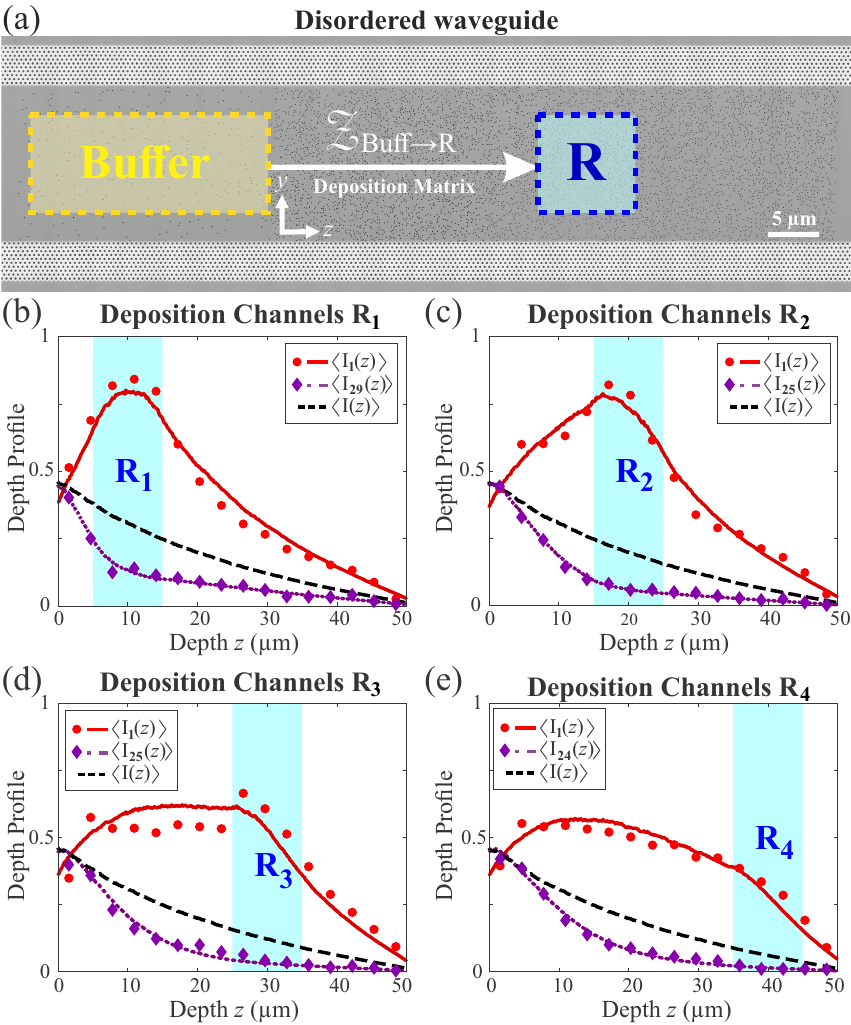}
\caption{ \label{fig:SupMeasure}
{\bf Experimental measurement of deposition eigenchannels.} A composite SEM image of an on-chip disordered waveguide is presented in (a) with a delineation of the buffer region and a deposition region superimposed. In (b-e) the maximally enhancing/suppressing (red-dots/purple-diamonds) deposition eigenchannel profiles, measured experimentally, are juxtaposed with numerical simulated profiles (red-solid and purple-dashed lines): for four target regions centered at 10, 20, 30, and 40 \textmu m.
} 
\end{center}
\end{figure}

In Figures~\ref{fig:SupMeasure} (b-e), the maximal enhancing/suppressing energy  deposition eigenchannel profiles (red-dots/purple-diamonds), measured experimentally, are juxtaposed with numerically simulated profiles (red-solid and purple-dashed lines): for all four target regions centered at 10, 20, 30 and 40 \textmu m. The cross-section integrated intensities are averaged axially (along $z$) over one transport mean free path to reduce fluctuations. The black dashed line represents the cross-section integrated intensity profile of random illumination patterns in our system, $\langle I(z) \rangle$. The depth profiles of the maximal energy deposition eigenchannels (with the largest eigenvalues) are well above $\langle I(z) \rangle$: both within the target region and beyond. Similarly, the depth profiles of the minimal energy deposition eigenchannels (with the smallest eigenvalues) are notably lower than $\langle I(z) \rangle$, both inside and outside the target region.

\newpage

\newpage

\newpage

\newpage

\newpage

\newpage
\end{document}